%% file: acena_MCMCI.tex
\documentclass[useAMS,usenatbib]{mn2e}

\usepackage{mathrsfs}
\usepackage{amsmath,amsfonts}
\usepackage[dvips]{graphicx}
\usepackage{xcolor}
\usepackage{rotating}
\usepackage{txfonts}
\usepackage{dsfont}
\usepackage[T1]{fontenc}
\usepackage{soul}
\usepackage{algorithmic}
\usepackage{algorithm}

\usepackage[multiple]{footmisc}
\usepackage{booktabs}
\usepackage{rotating}

\input{def_stars}
\input{def_units}
\input{def_symbs}

\def\thetav{\boldsymbol{\theta}}

\def\Xv{\mathbf{X}}
\def\stage{t_{\star}}
\def\ass{\langle \delta \nu \rangle}
\def\als{\langle \Delta \nu \rangle}
\newcommand{\run}[1]{run~\#{#1}}
\usepackage{color}
\definecolor{colres}{rgb}{0.996,0.945,0.813}

\def\epsilonb{{\boldsymbol{\epsilon}}}
\def\ind{\mathds{1}}
\def\Exp{\mathds{E}}
\def\pspace{\mathds{P}}
\def\ospace{\mathds{O}}
\def\ecno{\varepsilon_{\mathrm{CNO}}}
\def\ecnoc{\varepsilon_{\mathrm{CNO,c}}}
\def\xcno{X_{\mathrm{CNO}}}
\def\xcnoc{X_{\mathrm{CNO,c}}}
\def\ec{\varepsilon_{\textnormal{c}}}
\def\epp{\varepsilon_{\mathrm{pp}}}
\def\eppc{\varepsilon_{\mathrm{pp,c}}}
\def\tc{T_{\mathrm{c}}}
\def\xc{X_{\mathrm{c}}}

\def\grad{\nabla_{\mathrm{rad}}}
\def\mcc{m_{\mathrm{cc}}}

\title[]{A Bayesian approach to the modelling of $\boldsymbol{\alpha}$ Cen A}
\author[M.~Bazot, S.~Bourguignon and J.~Christensen-Dalsgaard]{M.~Bazot$^{1}$\thanks{E-mail: bazot@astro.up.pt}, S.~Bourguignon$^{2}$ and J.~Christensen-Dalsgaard$^{3}$\\
$^{1}$Centro de Astrof\'{\i}sica da Universidade do Porto, Rua das Estrelas, 4150-762, Porto, Portugal\\
$^{2}$Institut de Recherche en Communication et Cybern\'etique de Nantes, 1 rue de la No\"e, BP 92101, 44321 Nantes Cedex 3, France\\
$^{3}$Stellar Astrophysics Centre, Department of Physics and Astronomy, Aarhus University, Ny Munkegade 120, DK-8000 Aarhus C, Denmark}

\usepackage[pagewise,modulo,displaymath]{lineno}
\linenumbers

\begin{document}

\date{Accepted 2011 December 15. Received 2011 December 14; in original form 2011 October 11}

\pagerange{\pageref{firstpage}--\pageref{lastpage}} \pubyear{2002}

\maketitle

\label{firstpage}

\begin{abstract}
  Determining the physical characteristics of a star is an inverse problem consisting in estimating the parameters of models for the stellar structure and evolution, knowing certain observable quantities. We use a Bayesian approach to solve this problem for {\acena}, which allows us to incorporate prior information on the parameters to be estimated, in order to better constrain the problem. Our strategy is based on the use of a Markov Chain Monte Carlo (MCMC) algorithm to estimate the posterior probability densities of the stellar parameters: mass, age, initial chemical composition,\dots We use the stellar evolutionary code ASTEC to model the star. To constrain this model both seismic and non-seismic observations were considered. Several different strategies were tested to fit these values, either using two or five free parameters in ASTEC. We are thus able to show evidence that MCMC methods become efficient with respect to more classical grid-based strategies when the number of parameters increases. The results of our MCMC algorithm allow us to derive estimates for the stellar parameters and robust uncertainties thanks to the statistical analysis of the posterior probability densities. We are also able to compute odds for the presence of a convective core in {\acena}. When using core-sensitive seismic observational constraints, these can raise above $\sim$40\%. The comparison of results to previous studies also indicates that these seismic constraints are of critical importance for our knowledge of the structure of this star.  
\end{abstract}


\section{Introduction}
\input{Sections/Intro.tex}
\section{The Bayesian approach}\label{sect:Bayes}
\input{Sections/Subsections/Inverse.tex}
\input{Sections/Subsections/Proba.tex}

\input{Sections/Subsections/MCMC.tex}

\section{Stellar models}\label{sect:model}
\input{Sections/Subsections/Model.tex}
\section{Observational constraints}\label{sect:obs}
\input{Sections/Subsections/Observations.tex}

\section{Results on {\acena}}\label{sect:results}
\input{Sections/Subsections/2D.tex}

\input{Sections/Subsections/5D.tex}

%\section{Discussion}\label{sect:blabla}
%\input{Sections/Discussion.tex}
%\subsection{Impact of the seismic parameters}
%\subsection{Impact of the prior information}
%\subsection{Effect of alternative physics}
%\subsection{Comparison with previous works}
%\subsection{Strategies for stochastic sampling}\label{sect:MCMCimprov}
\section{Conclusion}
\input{Sections/Conclusion.tex}

\section*{Acknowledgments}
Micha\"el Bazot would like to thank S.~Hannestad for providing him access to the Grendel cluster at DCSC/AU of which important use has been made during this work. He would also like to thank H.~Carfantan for his kind support and J.~P.~Marques for very interesting and useful discussions. This work was co-supported by grants SFRH/BPD/47994/2008 and PTDC/CTE-AST/098754/2008 from FCT/MCTES and FEDER, Portugal. Funding for the Stellar Astrophysics Centre is provided by The Danish National Research Foundation. The research is supported by the ASTERISK project (ASTERoseismic Investigations with SONG and Kepler) funded by the European Research Council (Grant agreement no.: 267864). Finally, the authors would like to thank the anonymous referee for is work.

\bibliography{MCMCref}

\appendix
%\section{Computational performance of ASTEC}\label{app:CPU}
\section{Monitoring of convergence}\label{app:conv}
\input{Appendices/Conv.tex}
%\section{Tests on simulations}\label{app:simus}
%\input{Appendices/Simu.tex}

\end{document}

%% file: def_stars.tex
\def\acena{${\alpha}$~{Cen}~A}
\def\acenb{${\alpha}$~{Cen}~B}

\def\18sco{18 {Sco}}

%% file: def_units.tex
\def\m2s2{m$^{2}$~s$^{-2}$} %m2.s -2
\def\msol{M${_\odot}$}             %Msun
\def\rsol{R${_\odot}$}             %Rsun

%% file: def_symbs.tex
\def\teff{T_{\mathrm{eff}}}             %Effective Temperature

%% file: Sections/Intro.tex
In stellar physics, {\acena} holds a particular place. Located at 1.3~pc from the Sun, it is the closest star to the Solar System. Its apparent magnitude is $V = -0.01$, making it a privileged target to obtain precise and accurate data from spectroscopy, photometry, astrometry or interferometry. Having a G2V spectral type, it is physically close to the Sun and is thus of importance for physicists interested in the processes and mechanisms governing the structure and evolution of solar analogs. Finally, it is part of a triple system and forms a close, visual, binary with {\acenb}. This allows to directly measure the mass of the star, which is an information of critical importance in stellar physics.

For all these reasons, {\acena} has been the focus of many observational campaigns and theoretical studies. The former have allowed to obtain very precise data for the star. We possess measurements of its atmospheric parameters \citep[effective temperature, surface gravity, chemical abundances;][]{Chmielewski92,Neuforge97}, luminosity \citep{Soderhjelm99}, radius \citep{Kervella03}, oscillation frequencies \citep{Bouchy02,Bedding04},\dots All these have been combined to constrain theoretical models for the stellar structure and evolution, leading to a significant improvement of our knowledge of the physical properties of {\acena}, its mass, age, initial chemical composition.

Despite all these advances, it is obvious that there is no perfectly unified picture yet of the internal structure of {\acena}, neither of its evolutionary state. Even though its mass is well-known, there exist discrepancies in the ages found in the literature \citep{Turcotte98,Guenther00,Morel00,Eggenberger04,Miglio05}. It indeed depends very much on the considered observational constraints and the particular physics included in the models that are used to reproduce these quantities.

Furthermore, as it turned out, {\acena} is in a very peculiar state with regard to stellar structure and evolution. Indeed, stars with mass $\gtrsim$1.1~{\msol} can be convectively unstable in their cores. This is very important because it affects the energy transport in the star and thus its entire evolution. However, it is a difficult task to identify such small convective cores. First, they might not influence much the stellar observables (with the notable exception of its oscillation frequencies) and could remain undetected. Second, the physics of convection governing these structures is poorly known, and it is likely that the models used to describe {\acena} are incorrect with respect to the treatment of its inner convection. This is reflected by the fact that various studies have led to different claims regarding the existence of a convective core in {\acena} \citep{Thevenin02,Thoul03,Eggenberger04,Miglio05,deMeulenaer10}.

A further problem appears when one needs to quantify precisely the uncertainties on the physical stellar parameters estimated relatively to the uncertainties on the observations. This is not a problem specific to {\acena}: it has always been challenging in stellar physics to obtain robust and credible uncertainties on the parameters estimated from modelling. This is mostly due to the substantial computational cost of the stellar evolutionary codes, which is often incompatible with the intensive use of the models required by statistical methodologies. It does not come as a surprise then that most of the efforts in stellar physics have long been directed towards improving the codes themselves, either numerically or by including more accurate physics, rather than developing tools for parameter estimation.

This situation has progressively changed during the past decades. Many studies have been published providing extensive grids of models, spanning large ranges of values of physical parameters such as the stellar age, mass or initial chemical composition \citep{Schaller96,Girardi00,Pietrinferni04,Quirion10}. These can help to proceed to statistical studies without having to compute too many models at once. Simultaneously, there has been a growing interest for methodologies for parameter estimation, which have been applied to stellar physics. Among them, optimization methods have been introduced first, in some cases great care have been used to combine this procedure with uncertainty analysis such as SVD decomposition \citep{Brown94}. The methods adopted for optimization are many, from straightforward grid analysis to solving the non-linear least-square problem using a Levenberg-Marquardt algorithm \citep{Miglio05} or stochastic sampling via genetic algorithms \citep{Metcalfe03}. Other studies have focused on the application of Bayesian statistics with some success. Here again, the methodologies differ, some using grid-based strategies \citep{Jorgensen05,daSilva06,Takeda07} other Markov Chain Monte Carlo (MCMC) algorithms \citep{Bazot08}.

In this article, we present a Bayesian modelling of {\acena} based on such an MCMC methodology. In Sect.~\ref{sect:Bayes}, defining our inverse problem, we briefly recall the essentials of the Bayesian approach. We also give a brief account on the MCMC theory and present the algorithm we used to solve this inverse problem. In Sect.~\ref{sect:model} and \ref{sect:obs} we present the physical context to which these tools are to be applied. We review briefly the characteristics of our stellar evolution code and give an overview of the current observational knowledge on {\acena}. In order to assess the potential gains one can expect from the use of MCMC algorithms with respect to grid-based strategies we give two solutions to our inverse problem. In the first one we let two stellar parameters vary and make assumptions to fix three others. In the second one, we estimate all five. We report the results from both numerical experiments in Sect.~\ref{sect:results}. In particular, we concentrate on two points. One is physical: what is the general picture of the structural and evolutionary state of {\acena}? The other is numerical: how efficiently were these results obtained? From the methodological point of view, our results show the interest of MCMC approaches for such estimation problems, when no explicit mapping exists between observables and the parameters to be estimated. In particular, compared to grid-based computations, MCMC reveals better efficiency by concentrating most of the computational effort on the regions of interest. They also naturally provide crucial information about uncertainties associated to the estimated values. We briefly compare our results to previous studies in Sect.~\ref{sect:previous}

%% file: Sections/Subsections/Inverse.tex
\subsection{A difficult inverse problem}
\label{sec:pbinv}

Stellar models include several parameters that have to be estimated by matching observational data, which represents a typical inference problem. Stellar observable quantities usually considered are the luminosity $L$, the effective temperature $T_{\mathrm{eff}}$, the surface metallicity  $Z$, the radius $R$ and the frequencies of stellar pulsation modes in a given range of radial orders $n$ and angular degrees $l$, say $(\nu_{n,l})$. 
 {\acena} is one of the very important cases in stellar physics for which all of these measurements are available to high precision. 
 From such observations, one usually wants to estimate stellar parameters such as the mass of the star, $M$, its age, $\stage$,  the mixing-length parameter, $\alpha$, and the initial mass-fractions of helium and heavy elements, respectively $Y_0$ and $Z_0$.\footnote{Or $X_0$, the initial hydrogen fraction, remembering that the relation $X_0+Y_0+Z_0 = 1$ holds.}  This is by no mean an exhaustive set of parameters, in particular if non-standard physics are to be included.

 Unfortunately, there is no closed-form equation linking stellar parameters together with observables, which could be {\it inverted}. For a given set of interest stellar parameters, corresponding values of $L$, $T_{\mathrm{eff}}$, $Z$,  $R$ and $(\nu_{n,l})$ can be computed by stellar evolution codes (which may also require additional parameters). Let $\thetav = (\theta_1,\dots,\theta_K)$ collect the unknown stellar parameters and $\Xv = (X_1,\dots,X_N)$ collect the observed quantities. We shall call the parameter space, ${\pspace}$, and the observable space, $\ospace$, the spaces containing respectively $\thetav$ and $\Xv$. We have $\dim(\pspace) = K$. 

We shall now consider a mapping from $\pspace$ to $\ospace$ and write
\begin{equation*}
\widetilde{\Xv} = \pmb{\mathscr{S}}(\thetav) ,
\end{equation*} 
where elements in $\widetilde{\Xv}$ are the values predicted by the stellar code for these quantities, denoted here by $\pmb{\mathscr{S}}$. We consider that the observations are related to $\widetilde{\Xv}$ by
\begin{equation*}
\Xv = \widetilde{\Xv} + \epsilonb,
\end{equation*}
where $\epsilonb$ is an unknown error term accounting for instrumental errors on the measurements and possible model errors between $\widetilde{\Xv}$ and $\pmb{\mathscr{S}}(\thetav)$. The problem addressed here concerns the estimation of $\thetav$ from
\begin{equation}
\Xv =\pmb{\mathscr{S}}(\thetav) + \epsilonb,
\label{eq:pbinv}
\end{equation}
which defines an {\it inverse problem}~\citep{Idier08}. 
A classical approach to solve this problem consists in minimizing an appropriate distance, in $\ospace$, between $\Xv$ and $\widetilde{\Xv}$, such as the usual $\chi^2$ error $\|{\Xv - \pmb{\mathscr{S}}(\thetav)\|}^2 = \sum_{n=1}^N (X_n - \widetilde{X}_n)^2$. This is a hard task in our case. First, $\pmb{\mathscr{S}}(\thetav)$ is not an explicit function, so no analytical formulation can be exploited by efficient optimization methods (for example, no expression of the derivatives of $\pmb{\mathscr{S}}(\thetav)$ with respect to each variable in $\thetav$ can be used for optimization). Second, stellar models computed by $\pmb{\mathscr{S}}$ make use of highly non-linear functions, so that classical optimization methods may be stacked in local minima or yield degenerate solutions. The complexity of the mapping from the space of parameters to the space of observables has already been pointed out \citep{Brown94,Jorgensen05} and represents one of the major challenges in stellar-parameter estimation. It has to be noted that obtaining $\pmb{\mathscr{S}}(\thetav)$ for a given set of parameters $\thetav$ is generally computationally expensive. Therefore, computing the model on a grid of discrete values of the parameters may be extremely time-consuming if one wants to explore the whole parameter space with high precision, especially as its dimension increases. 

We consider the use of Markov Chain Monte Carlo methods~(MCMC) as an alternative to classical optimization methods, where model computations are concentrated around the interest regions of the parameter space, and are in that sense "optimized". Basically, the distance between observations and their prediction by $\pmb{\mathscr{S}}(\thetav)$ is linked to a probability distribution, and MCMC algorithms aim at approximating it by generating random samples distributed according to such distribution. Then, they allow one to obtain not only an estimate of the interest parameters, but also to associate characterizations such as uncertainties or higher-dimensional statistics. MCMC are also naturally oriented toward a Bayesian setting, in which prior information on the unknown parameters is also taken into account by means of probability distributions. Furthermore, being stochastic in nature, they might help to avoid confinement in local minima.

%% file: Sections/Subsections/Proba.tex
\subsection{Probabilistic framework}

\subsubsection{Likelihood function}

Equation~(\ref{eq:pbinv}) has a statistical interpretation, where the  perturbation term $\epsilonb$ can be seen as the realization of a random process. We suppose hereafter that the error $\epsilon_n$ on each observation $X_n$ is normally distributed, with  variance $\sigma_n^2$ and that the errors affecting the different observations are independent one from each other. 
In our case, we assume that $\sigma_n$ corresponds to the uncertainty on $X_n$, which is determined by the observation. 

Then, the probability density of all perturbation terms taken together is $$p_\epsilonb(\epsilonb) = \prod_{n=1}^N \frac{1}{{(2\pi)}^{1/2}\sigma_n } \exp \left(-\frac{\epsilon_n^2}{2\sigma_n^2}\right),$$ which  defines the {\it likelihood function} of data $\Xv$ given a model parameterized by $\thetav$, using (\ref{eq:pbinv})
\begin{eqnarray}
\mathscr{L} (\thetav;\Xv )& = &p_\epsilonb \left(\Xv -\pmb{\mathscr{S}}(\thetav)\right)  \nonumber\\
& = &  \displaystyle\prod_{n=1}^{N} \frac{1}{(2\pi)^{1/2}\sigma_n}  \dots \nonumber\\
&&\times \exp \left( -  \displaystyle \frac12 \sum_{n=1}^{N} \left[ \frac{\mathscr{S}_n(\thetav)-X_n}{ \sigma_n} \right]^2 \right).
{\label{eq:likeli}}
\end{eqnarray}
Then, maximizing the likelihood~(\ref{eq:likeli}) is equivalent to minimizing the usual least-squares term
\begin{equation}
\min_{\thetav} \chi^2(\thetav) = \sum_{n=1}^{N} \left[ \frac{\mathscr{S}_n(\thetav)-X_n}{ \sigma_n} \right]^2.
\label{eq:minchi2}
\end{equation}

\subsubsection{Incorporating prior information}

Suppose one has some --~even weak~-- prior information about the unknown parameters $\thetav$, such as an acceptable range of possible values or previous measurements with associated uncertainties. 
One may want to take it into account in order to constrain the solution of~(\ref{eq:minchi2}). Bayesian statistics provide a natural way to do this~\citep{Idier08}, by considering such information through a prior probability distribution on~$\thetav$, say~$\pi(\thetav)$. 

Suppose, for example, that the age $\stage$ is bounded between $t_{\text{min}}$ and $t_{\text{max}}$, without more information. Then one can use 
\begin{equation}\label{eq:uprior}
\pi(\stage) = \frac{1}{t_{\text{max}} - t_{\text{min}}}  \ind_{[t_{\text{min}},t_{\text{max}}]}(\stage),
\end{equation}
where $\ind_{\mathscr{D}}(u)$ is the indicator function of the domain $\mathscr{D}$, which equals $1$ if $u \in \mathscr{D}$ and $0$ otherwise. That is, $\pi(\stage)$ is a uniform distribution on $\mathscr{D}$.
Or suppose that the mass $M$ has been previously measured at $M_{\text{est}}$, with uncertainty $\sigma_{M_{\text{est}}}$. Then, one may use a Gaussian prior distribution $\pi(M)$, centred at $M_{\text{est}}$ and with standard deviation $\sigma_{M_{\text{est}}}$ (or $2\sigma_{M_{\text{est}}}$, $3\sigma_{M_{\text{est}}}$ for a weaker prior distribution). 

Once such priors are defined, $\thetav$ can be estimated from the posterior probability distribution~(PPD), which reads, by Bayes' theorem
\begin{equation}{\label{eq:bayes}}
\pi(\thetav | \Xv)=\frac{\mathscr{L}(\thetav; \Xv)\pi(\thetav)}{m(\Xv)}.
\end{equation}
$\pi(\thetav | \Xv)$ represents the probability distribution of all unknown parameters constrained by observed values of~$\Xv$ and by prior distributions. 
In the latter equation, $m(\Xv)$ is a normalization constant ensuring that the integral of $\pi(\thetav | \Xv)$ over the entire space of parameters is 1.

\subsubsection{Exploiting the posterior distribution}

Suppose we are able to access the whole posterior probability distribution~$\pi(\thetav|\Xv)$. It contains all statistical information available, given a set of related observed quantities and (possibly) under the given prior assumptions. If no prior distribution is used, then $\pi(\thetav|\Xv)$ identifies with the likelihood function~$\mathscr{L}( \thetav; \Xv)$.
We are then theoretically able to exploit such a distribution in order to derive estimates of interest parameters~$\thetav$.  In this paper, we  focus on the two most common ones in Bayesian analysis. The most intuitive one is the Maximum A Posteriori (MAP) estimate, which is the set of parameters that maximize the PPD (that is, "the most probable value" of $\thetav$ given observations $\Xv$)
\begin{equation}\label{eq:map}
\widehat{\thetav}_{\text{MAP}} = \arg\max_{\thetav} \pi(\thetav | \Xv).
\end{equation}
The Posterior Mean (PM) estimate corresponds to the expected value of $\thetav$ under distribution (\ref{eq:bayes})
\begin{equation}\label{eq:pm}
\widehat{\thetav}_{\text{PM}} = \Exp[\thetav|\Xv] = \int \thetav \pi(\thetav | \Xv) d\thetav.
\end{equation}
It minimizes the mean square error $\Exp\left[(\widehat{\thetav}(\Xv) - \thetav)^2\right]$.

The latter one is also often used in Bayesian estimation since, knowing $\pi(\thetav | \Xv)$, one can associate variances, that is, uncertainties, to the estimated values.

In practice, of course, we do not have access to the whole posterior probability distribution and computation of the former estimates is a hard task. Optimization of~$\pi(\thetav|\Xv)$ is at least as difficult as optimization of likelihood~(\ref{eq:likeli}), for the reasons explained in Sect.~\ref{sec:pbinv}: $\pi(\thetav | \Xv)$ does not have an exploitable closed-form expression. Computing $\widehat{\thetav}_{\text{PM}}$ requires an integration over all possible values of~$\thetav$ and is also impossible analytically.

%% file: Sections/Subsections/MCMC.tex
\subsection{Markov Chain Monte-Carlo algorithm}

MCMC methods~\citep[for a general review see][]{Robert05} are powerful techniques allowing to approximate complex probability densities. Roughly speaking, MCMC aims at generating samples randomly distributed according to the interest distribution (\ref{eq:bayes}), and then approximating such distributions by histograms.  
MCMC methods are powerful in the sense that they are able to generate random samples distributed (at least, asymptotically) according to a target distribution only by evaluating this distribution at a certain number of parameter values.

In our case, suppose we have $T$ samples~$\thetav^{(1)},\dots,\thetav^{(t)},\dots,\thetav^{(T)}$ distributed according to $\pi(\thetav|\Xv)$ -- we use the notation $\thetav^{(t)} \sim \pi(\thetav|\Xv)$. Then, the PM estimate~(\ref{eq:pm}) of all parameters in~$\thetav$ can be approximated by
\begin{equation}
\widehat{\thetav}_{\text{PM}} \simeq \frac{1}{T}\sum_{t=1}^T \thetav^{(t)},
\end{equation}
and the corresponding unbiased variance estimator by
\begin{equation}
\widehat{\sigma}^2\left(\widehat{\theta}_{k,\textnormal{PM}}\right) \simeq \frac{1}{T-1}\sum_{t=1}^T \left(\theta_k^{(t)} -\widehat{\theta}_{k,\textnormal{PM}}\right)^2,
\end{equation}
with $\theta_k$ the $k$-th component of $\thetav$. We can also approximate the MAP estimate by taking, among all~$\thetav^{(t)}$, the set of parameters for which~$\pi(\thetav^{(t)}|\Xv)$ is the highest (note that the corresponding values are naturally obtained as a by-product of the MCMC algorithm). 

MCMC methods work by generating a {\it Markov Chain}, that is, a sequence $\thetav^{(1)},\dots,\thetav^{(T)}$ such that the rule for generating~$\thetav^{(t)}$ depends only on the previous value~$\thetav^{(t-1)}$. The {\it Monte-Carlo} reference obviously concerns the random nature of the process. 
Giving a full account of the theory of MCMC is obviously far beyond the scope of this paper. In the following, we will only comment on some essential practical aspects necessary for the comprehension of the subsequent parameter estimation. This approach was first suggested in the framework of stellar modelling by \citet{Bazot08}

\subsubsection{The Metropolis-Hastings algorithm}

In this paper, we construct a Metropolis-Hastings (MH) algorithm \citep{Metropolis53,Hastings70}, which is the most generic class of MCMC methods. The generality of this method allows to estimate probability densities of all forms, with the only assumption that they can be evaluated at every point in the parameter space. The PPD expressed in~(\ref{eq:bayes}) falls within this category. 

The Markov Chain is constructed through a two-step iterative procedure.  Suppose that $\thetav^{(1)},\dots,\thetav^{(T)}$ have been obtained.  First, a trial value of the set of parameters, $\thetav^{\ast}$, is generated according to an {\it instrumental distribution} $q(\thetav^{\ast}|\thetav^{(t)})$, that may depend on the previous value of the chain~$\thetav^{(t)}$.  Then a selection process operates, where~$\thetav^{(t+1)}$ is set to $\thetav^{\ast}$ with some probability~$\rho$, and to $\thetav^{(t)}$ otherwise. The convergence of the distribution of~$\{\thetav^{(t)}\}$ toward the target distribution is then ensured by the analytical expression  of~$\rho$, see the algorithm description below. We give more details about monitoring of the convergence of MCMC in Appendix~\ref{app:conv}.

{\usefont{T1}{cmvtt}{m}{n}

\begin{algorithm}[h!]
\caption{Metropolis-Hastings algorithm}
\label{algo:MHalg}
For $t=1$, choose an initial value~$\thetav^{(1)}$. Then, at iteration $t$:
\begin{algorithmic}
\STATE 1. Generate: $\thetav^{\ast} \sim q(\thetav |\thetav^{(t)})$
\STATE 2. Select:  \begin{equation*}\thetav^{(t+1)} = \begin{cases}
\thetav^{\ast} &\text{with probability $\rho(\thetav^{(t)},\thetav^{\ast})$,}\\
\thetav^{(t)} &\text{with probability $1-\rho(\thetav^{(t)},\thetav^{\ast})$,}
\end{cases}
\end{equation*}
\STATE  \ \ \ \ where \begin{equation*}
\rho(\thetav^{(t)},\thetav^{\ast})= \min  \left\{ \frac{\pi(\thetav^{\ast}|\Xv)}{\pi(\thetav^{(t)}|\Xv)} \frac{q(\thetav^{(t)}|\thetav^{\ast})}{q(\thetav^{\ast}|\thetav^{(t)})} , 1\right\}.
\end{equation*}
\end{algorithmic}
\end{algorithm}
}
The choice of $\thetav^{\ast}$ at step 1  is detailed in~\S~\ref{par_instrumental}. In practice, accepting $\thetav^{(t+1)} =  \thetav^{\ast}$ with probability~$\rho$ is performed by generating~$u$ according to a uniform distribution in~$[0,1]$, and  $\thetav^{(t+1)} =  \thetav^{\ast}$ is accepted if $u < \rho$.

Note that if $q$ is a symmetrical distribution (that is, $q(\thetav^{(t)}|\thetav^{\ast}) = q(\thetav^{\ast}|\thetav^{(t)})$), which will be the case here, then the ratio in probability~$\rho$ is nothing but the PPD ratio evaluated at $\thetav^{\ast}$ and $\thetav^{(t)}$. Hence, parameters increasing the PPD are always accepted while parameters decreasing the PPD are randomly accepted, the  probability of acceptance decreasing with the PPD ratio.

\subsubsection{Instrumental distribution}
\label{par_instrumental}

A key point conditioning the efficiency of the MH algorithm is the choice of the instrumental distribution. First, it should be computationally easy to draw random samples distributed according to $q$. Second, $q$ should concentrate the proposed values~$\thetav^{\ast}$ around the interest regions of~$\pi(\thetav|\Xv)$. Finally, $q$  must also explore the whole support of possible values in order not to miss plausible parameter sets. It should be noted that the first condition represents one of the major computational interests of using MCMC algorithms: it becomes possible to simulate complex distributions using only simpler ones.

Trying to account for these issues and following former similar choices~\citep[e.g.,][]{Andrieu99} we choose~$q(\thetav^{\ast} | \thetav^{(t)})$ as a mixture of two Gaussian distributions centred at $\thetav^{(t)}$ and the uniform distribution on the definition domain of all parameters. That is, for Step 1 in the MH algorithm, we generate, for all parameters~$\theta^{\ast}_k$, with $k=1,\dots,K$

\begin{equation}{\label{eq:inst}}
\begin{cases}
\theta^{\ast}_k \sim {\mathscr{N}}_{\mathscr{D}_{\theta_k}} \left(\theta_k^{(t)},\varsigma_{k,1}^2\right) &\text{ with probability } p_1,   \\
\theta^{\ast}_k \sim {\mathscr{N}}_{\mathscr{D}_{\theta_k}} \left(\theta_k^{(t)},\varsigma_{k,2}^2\right)&\text{ with probability } p_2, \\
\theta^{\ast}_k \sim {\mathscr{U}}_{\mathscr{D}_{\theta_k}}&\text{ with probability } p_3.
\end{cases}
\end{equation}
where ${\mathscr{N}}_{\mathscr{D}_{\theta_k}} \left(\theta_k^{(t)},\varsigma_{k,1}^2\right)$ is the Gaussian distribution with mean $\theta_k^{(t)}$ and variance $\varsigma_{k,1}^2$, truncated to the definition domain~${\mathscr{D}_{\theta_k}}$ of~$\theta_k$, and ${\mathscr{U}}_{{\mathscr{D}_{\theta_k}}}$ is the uniform distribution on~${\mathscr{D}_{\theta_k}}$ --~see Sects.~\ref{sect:simu2D} and \ref{sect:simu5D} in which we specify the $(\varsigma_{k,j})$, $p_i$ and the acceptable definition domain for~$\thetav$.

 The use of Gaussian distributions centred on the current value  $\thetav^{(t)}$ defines random walks in the space of parameters. Having two of them helps to perform local scans of the parameter space at different scales. We added the uniform distribution to avoid possible trapping in local extrema: if the chain is being stuck in such an area, there is chance for the algorithm to propose, from time to time, values of $\thetav$ well outside this region.

%% file: Sections/Subsections/Model.tex
\subsection{Physics}{\label{sect:stmod}}

In Eq.~\ref{eq:pbinv} appears the quantity $\pmb{\mathscr{S}}(\thetav)$, describing the predicted values for the observable quantities. These values were computed using the Aarhus STellar Evolution Code (ASTEC). A thorough description is given by \citet{JCD82b,JCD08a}, we thus only briefly recall the main outline of the physics used in our models. The oscillation frequencies were computed with the {\tt adipls} package\citep{JCD08b}. We restrict ourselves to the so-called standard physics. This implies a spherical, non-rotating star. Microscopic diffusion, magnetic field and mixing processes other than convection (overshooting, rotational mixing,...) are neglected. This latter is described using the mixing-length theory \citep{BV58} that requires the setting of a mixing-length parameter, $\alpha$, defined as $\alpha \equiv \ell_m/H_p$, where $\ell_m$ is the mixing-length and $H_p$ is the pressure scale-height..

The physics used in the stellar code is relatively basic. It is not the goal of this study to adopt the state-of-the-art physics for the stellar evolution code. The emphasis was rather put on the need to use a fast, robust and already tested version of the code. This is the reason why we used the EFF tables for the equation of state \citep{EFF} instead of the latest OPAL tables \citep{OPAL96,OPAL02}, noting however that they agree {\textquotedblleft}reasonably well{\textquotedblright}. It also appears from \citet{Miglio05} that, in the framework of the modelling of {\acena}, using EFF does not give results departing significantly from those using OPAL. Other input physics included the NACRE nuclear reaction rates \citep{Angulo99} and OPAL opacity tables \citep{Iglesias96}. The relative abundances of heavy elements are those prescribed by \citet[GNS abundances]{Grevesse93}. %\hl{Travis' talk in Lanzarote for effect of EFF}

This latter point should be briefly commented since these abundances are, at the moment, subject to debate. Using 3-dimensional simulations, \citet{Asplund04} have derived new solar abundances (AGS abundances). The most remarkable features are the low values found for C, N, O. Abundances of other heavy elements such as Na, Mg or Fe were also revised to lower values. The overall impact is a reduction of $\sim30\%$ of the estimated solar metallicity. However, including these values in solar models has led to sound-speed profiles in disagreement with helioseismology \citep[e.g.][]{T-C04,Montalban04,Antia05}. Although several attempts have been made, some of them including non-standards physics \citep[e.g.,][]{Guzik05,Castro07}, no satisfactory agreement between theoretical models and seismology could be reached. On the contrary, models using the GNS abundances, computed using 1-dimensional models, can reproduce with a remarkable precision the solar sound-speed profiles. It should be noted that recent studies \citep[e.g.,][]{Caffau08,Asplund09} have led to upward revisions of these abundance. Discriminating between AGS and GNS abundances is beyond the scope of this study. We thus keep on using the GNS heavy-element abundances in our models of {\acena}. This rather conservative course of action is motivated by the fact that we focus on testing a new technique for statistical inference of stellar parameters. That should be done with input physics that has already led to coherent results. At the very least, the inferred parameters for {\acena} could be compared to other stellar models that also used GNS abundances.

Given these physical prescriptions, it is important to note that the {\textquotedblleft}natural{\textquotedblright} parameter space is such that $\dim(\pspace) = 5$, with the typical associated vector $\thetav=(M,\stage,\alpha,Z_0,X_0)$.

%% file: Sections/Subsections/Observations.tex
In the following we review briefly the values we selected for the different observables for {\acena}, their uncertainties and their potential bias. We present both the non-seismic and the seismic observations. Table~\ref{table:obsns} sums up the former ones. For the oscillation frequencies, the reader is referred to \citet{Bazot07}.
  
\subsection{Spectro-photometric observables}

\input{Tables/obsns.tex}

Several spectroscopic determinations of the effective temperature of {\acena} have been carried out, whether by fitting the wings of the H$_{\alpha}$ line or the FeI lines \citep{Neuforge97}. However some discrepancies appear between the value used in the modelling by \citet{Morel00}, $5790\pm30$~K obtained by fitting the wing of the H$_\alpha$ line, and the value given by \citet{Neuforge97}, $5830\pm30$~K. To solve this problem, \citet{Eggenberger04} have adopted an average value of these two measurements and 1$\sigma$ error bars encompassing their uncertainties; the same value was used by \citet{Miglio05}. Since it can help to compare our results with these works, we also use $T_{\mathrm{eff}}=5810\pm50$~K, even though reduced error bars of $\pm30$~K could have helped reducing the uncertainties on the estimated parameters.

Although the proximity of {\acena} should allow a precise measurement of its parallax, discrepancies also subsist between different studies, affecting the subsequent values of the luminosity. We select the parallax determined by \citet{Soderhjelm99}. The corresponding luminosity is $L/L_{\odot} = 1.522\pm0.030$ \citep{Eggenberger04}. Note that this value of the parallax has also been used to derive the radius and the mass we describe below.

\subsection{Radius}{\label{radcon}}

The radius of {\acena} has been measured by \citet{Kervella03} with VINCI, the VLTI Commissioning Instrument, which operates in the K-band (2.0-2.4 $\mu$m). They measured an angular diameter, after correcting for limb-darkening effects, $d = 8.511 \pm 0.020$ mas. Combined with the parallax derived by \citet{Soderhjelm99}, it led to $R=1.224 \pm 0.003$ {\rsol}. For the reference solar radius value, we used the value measured from helioseismology by \citet{Brown98}, $R_{\odot} = 695.508$~Mm. It is noteworthy that the authors argued that this value, measured in the near-IR, can be compared to models computed with the stellar evolution code CESAM \citep{Morel97} since the radius is there defined as the point where $T=T_{\mathrm{eff}}$, which is close to the temperature of the layer at which the continuum at 2.2 $\mu$m is formed. Since ASTEC uses the same atmospheric boundary conditions \citep{JCD82b}, we consider that this assumption still holds in our case. 

 This radius value has been used in \citet{Eggenberger04}, along with seismic data from \citet{Bouchy02}, to constrain their model. They concluded that they cannot reproduce at the same time the seismic measurements and the radius within their $1\sigma$ error bars. However, they noted that this could be done when considering a $2\sigma$ interval. This confirmed the $1.3\sigma$ agreement \citet{Kervella03} found between the observed value of the radius and a previous theoretical estimate using CORALIE frequencies \citep{Thevenin02}. \citet{Miglio05} did not use the observed radius as a constraint for their models but always obtained values higher than the measured one. \citet{Montalban06}, using this constraint in combination with seismic data from \citet{Bedding04}, were able to reproduce all observations within $1\sigma$.

\subsection{Seismic observations}

\subsubsection{Individual frequencies}
The frequencies used in this paper are those from \citet{Bazot07}. As far as their accuracy is concerned, they can be considered as reliable. The high SN ratio that can be achieved in the radial-velocity spectrum of {\acena} allows to find easily the most prominent modes, whose amplitude can reach up to 14 times the average noise level. A comparison between the studies by \citet{Bouchy02,Bedding04} and \citet{Bazot07} offers a convincing empirical evidence that the modes are indeed real. The identifications presented in these studies are consistent (which of course does not prove them correct).

Of much greater difficulty is the task of estimating confidence intervals on the estimated frequencies. The assumption made in \citet{Bazot07} may be viewed as the crudest one: considering that the modes are unresolved, the frequency resolution is chosen as the uncertainty on the frequency. The generic uncertainty given, 1.3~$\mu$Hz, may be viewed as conservative. 
At the time our MCMC simulations were first performed, this was as good an estimate as one could find. We note however that methods are currently under development to allow better and more trustworthy estimation of the confidence intervals on the oscillation frequencies of solar-like stars \citep{Brewer07,Stahn08,Benomar10,Handberg11} A recent discussion on this methods has shown that they are efficient even for ground-based gaped and irregularly sampled data \citep{Bazot12}. While keeping in mind that room for potential improvement exists, we nevertheless used the original uncertainties in the present study, since their values may at the very least represent a realistic upper bound. We should also note that a new seismic study of {\acena} using the previous data of \citet{Bouchy02} and \citet{Bedding04} has been published by \citet{deMeulenaer10} during our study. They have not been included here, but we note that their result partially agree with those of \citet{Bazot07}.

Finally, we should note that fitting the individual frequencies is difficult because they are affected by surface effects that we cannot model. There have been tentatives to bypass this problem \citep{Kjeldsen08} and thus allow to use full sets of frequencies as an observational constraints. These methods have gained in popularity and are often used in the framework of Kepler for the modelling of stars. However, these strategies are empirical and we chose not to use them in the present study. We instead rely on the frequency combinations that we present below. A possible development to be tested along these lines would consist in including in our estimation the Bayesian formalism given by \citet{Gruberbauer12} that deals with this problem. This is left for future studies.

\subsubsection{Large separations}
The first asymptotic relation was given in the limit of low-degree, high-order p modes by \citet{Vandakurov67}. \citet{Tassoul80} further expanded the eigenfrequency equation to higher orders
\begin{equation}\label{eetassoul}
\begin{split}
\nu_{n,l} = (n + \frac{l}{2} +& \epsilon) \Delta \nu_0\\
&+ (l(l+1)A - B) \frac{\Delta \nu_0^2}{\nu} + \mathcal{O}(\frac{1}{\nu^2})
\end{split}
\end{equation}
with
\begin{equation}\label{eq:largesep}
\displaystyle \Delta\nu_0 = \left( 2\int^{R_t}_0 \frac{dr}{c} \right)^{-1},
\end{equation}
\begin{equation}\label{smallsep}
\displaystyle A=\frac{1}{4\pi^2\Delta \nu_0}\left[ \frac{c(R_t)}{R_t} -\int_{0}^{R_t} \frac{1}{r}\frac{dc}{dr}\frac{dr}{r}  \right],
\end{equation}
and with $R_t$ the upper turning point of the mode and $B$ a surface phase term.

If we assume, to the lowest order of approximation, that the phase-related term quantity $\epsilon_s$ does not depend on the frequency, we immediately see that modes of consecutive order and same degree are equally spaced. The difference $\Delta\nu_{n,l} = \nu_{n,l} - \nu_{n-1,l}$ is called the large separation and may be considered roughly as a constant. This has been a useful tool in helio-and asteroseismology to identify pulsation modes \citep[e.g.,][]{Grec83,Bouchy02}.

The large separation is, according to Eq.~(\ref{eq:largesep}), the inverse of twice the acoustic radius (i.e. the travelling time of an acoustic wave from the centre to the surface). We thus obtain a new global quantity that can be used to constrain stellar models. As noted by \citet{Deubner84}, this integral depends mostly on the external regions of the star. This can be easily seen if one considers that the sound speed in the stellar interior is a roughly monotonic, smoothly decreasing function of the radius, and that the ratio of its values at the centre and at the surface for a solar-like star is $\sim 70$. Therefore any acoustic wave will spend a much longer time closer to the surface than to the centre and will be predominantly sensitive to this region.

It has been been noted that the large separation depends on the frequency. Allowing for a the surface-phase term to vary with frequency, in an equation similar to (\ref{eetassoul}), \citet{Vorontsov91} pointed out that it will control the overall behaviour of $\Delta \nu_0$ when seen as a function of $\nu$. Once more, some problems arise when it comes to stellar models: as noted by \citet{Vorontsov89}, they usually provide poor values for this surface term and its frequency derivative, which is directly related to $\Delta \nu_0$.

From the point of view of stellar parameter estimation, this is nevertheless an interesting quantity. Besides the new (and relatively easy to extract) global constraint it provides, it is possible to compare the stellar parameters inferred using the large separations and other seismic indicators, obtained using different frequency combinations sensitive to deeper layers. This may in turn give us some insight on the modelling errors occurring below the surface. The quantity we used is the average large separation that we define as $\als = \Delta\nu_0$ and compute as explained in \citet{Bazot07} and note that the estimate they give should read $105.9\pm0.1$~$\mu$Hz and not $105.9\pm0.3$~$\mu$Hz; we therefore adopt the former value.

Numerically, we computed the average large separation by performing a linear regression, for each degree $l=0,1,2$, of the relation $\nu = \nu(n)$, only selecting in the code output the modes with orders corresponding to the values observed for {\acena}. This gives us three slopes, which are the average large separation for each degree. We then averaged over $l$ to obtain the average large separation.

\subsubsection{Small separations}
The second quantity used as a seismic indicator comes also straightforwardly from Eq.~(\ref{eetassoul}). It immediately appears that modes separated by one one order and two degrees, $\nu_{n,l}$ and $\nu_{n,-1,l+2}$, are degenerate at the first order. The difference between these frequencies is a second order quantity in $1/\nu$. It is called the small separation. 

Contrary to the large separations, the small ones depend mostly on the deep interior of the star. This can be seen from Eq.~(\ref{smallsep}). The integral term behaves as $c/r$ which is large at the centre and negligible near the surface. In-depth analysis by \citet{Gough86}, \citet{Gabriel89} and \citet{VanHoolst91} described much more rigorously this dependence.

Small separations are obviously a very important tool for stellar parameter estimation using seismology. It is noteworthy that \citet{JCD93} has shown that, in principle, the knowledge of the average values of the large and small separations is sufficient to characterize fully the evolutionary state of the star. We performed test estimation using the individual and the average large separations. The main difference is that the small uncertainty on the latter provides a better constraint to our stellar model, therefore, we adopted the value provided by \citet{Bazot07}, $\ass = 6.9\pm0.4$~$\mu$Hz. The average small separation in our models was simply computed by averaging our theoretical individual separations over the mode order (selecting only the observed ones).

Finally, we also tested the use of the ratio describe by \citet{RV94}, which should be completely free of surface effects\footnote{Whereas there is still a contribution of the surface layers to the small separations. In practice, our test runs with this so-called Roxburgh's ratio on {\acena} gave us the same estimates as those using the small separations, albeit with much larger error bars.}. However, the large uncertainty on these quantities, in the range 22.4\%-30.2\%, made them poor constraints in the case of this {\acena} data set.

\subsection{Using a mass measurement as a prior}{\label{sect:mprior}}

  {\acena} being part of a close binary, its mass can be measured using astrometry. Although its period is somewhat long, observations have been carried out throughout the last century, culminating with the precise measurements of \citet{Murdoch93}, \citet{Endl01} and \citet{Pourbaix02}, using high-precision spectrometers. This led to the derivation of a precise mass by \citet{Pourbaix02}, $M_{\mathrm{obs}} = 1.105 \pm 0.007$~{\msol} (the error being given to the $1\sigma$ level). This mass was also used by \citet{Eggenberger04}, \citet{Miglio05} and \citet{Montalban06}.

 We used this value to obtain prior information on the mass. The assumption is made that the uncertainties are Gaussian and reflect truly the 1$\sigma$ deviation from the mean value, therefore
\begin{equation}\label{eq:mprior}
\pi(M) \propto \exp\left( -\frac{(M-M_{\mathrm{obs}})^2}{\sigma^2_M} \right)
\end{equation}
It should be noted that \citet{Pourbaix02} issued a warning concerning the accuracy of this value. Even though the mass ratio between {\acena} and {\acenb} can be accurately constrained, the same is not true for their individual masses, because of the combined effects of the gravitational redshift and convective blueshift (this latter being especially difficult to correct for).

%% file: Tables/obsns.tex
\begin{table}
\begin{center}
\caption{Observational constraints on {\acena}. The last column specifies the reference for the selected quantity.}
\label{table:obsns}
\begin{tabular}{@{}lcr@{}}
\toprule
Observable& Value  & Reference\\
\midrule
 $T_{\mathrm{eff}}$& $5810\pm50$~K& \citet{Eggenberger04} \\
 $L/L_{\odot}$& $1.522\pm0.030$& \citet{Eggenberger04}\\
 $R/R_{\odot}$& $1.224\pm0.003$& \citet{Kervella03}\\
 $Z/X$& $0.039\pm0.006$& \citet{Thoul03}\\
 $\als$&$105.9\pm0.1$~$\mu$Hz&\citet{Bazot07}\\
 $\ass$&$6.9\pm0.4$~$\mu$Hz&\citet{Bazot07}\\
 $M/M_{\odot}$& $1.105\pm0.007$& \citet{Pourbaix02}\\
\bottomrule
\end{tabular}
\end{center}
\end{table}

%% file: Sections/Subsections/2D.tex
\input{Tables/simu2D_all.tex}

\input{Figures/PPD_Mt.tex}

\subsection{MCMC simulations with $\pmb{\dim(\pspace) = 2}$}\label{sect:simu2D}

\subsubsection{Implementation details}\label{sect:simu2D_implementation}

Our first set of MCMC simulations aimed at sampling the PPD in a parameter space of small dimension. This has some advantages, in particular for the comparison with procedures involving direct integration of the normalization factor in equation Eq.~(\ref{eq:bayes}) from fixed grids of models (Sect~\ref{sect:grids2D}). It is indeed possible, in that case, to compute very dense grids of stellar models. We thus limit ourselves to a parameter space such as $\dim(\pspace) = 2$, with the stellar mass and age the only two free quantities. 

In order to fix the other parameters, we had to make a certain number of assumptions. The initial hydrogen fraction was considered roughly solar  $X_0 = X_{0,\odot} \sim 0.7$. Combining with the [Fe/H] ratio for {\acena}, and because we neglected diffusion, this implies $Z_0 = 0.027$. This of course is an oversimplification since we intentionally neglect the effect of the uncertainty on [Fe/H]. However, such a configuration may happen if one is, for instance, using grids of stellar models only sparsly sampling the initial metallicity parameter. The mixing-length parameter is somewhat more problematic since there is no fundamental physical justification for introducing this parameter. It can be related to the gradient of entropy in the considered convective zone, but there is no physical prescription that can be unequivocally applied to this quantity. It is possible to use a canonical value, calibrated from the solar case. This is the approach we used here, setting $\alpha = \alpha_{\odot} \sim 1.8$. Some numerical simulations suggest that the value of  the mixing-length parameter in the solar neighbourhood in the $(M-\stage)$ plane is close to constant \citep{Trampedach07}.\\

The set-up was similar for all our MCMC simulations. For each combination, $\Xv$, of the observables we considered, we launched $m=3$ independent
Markov chains of length $T=50000$. The use of parallel chains is often a good method to control how dependent are the results on the initial guesses. The length of the chains was chosen, after some test runs, so that it was always longer than the effective number of iterations required to reasonably conclude that convergence has been reached. For each run, convergence was tested using graphical indicators such as the sequential mean and variance. A typical burn-in sequence of 5000 iterations was discarded at the beginning of each Markov Chain realization in order to ensure that the sample was indeed generated according to the target PPD (see Appendix~\ref{app:conv}).

Based on test runs, we set the Gaussian components of the instrumental law $q(\thetav^{\ast}|\thetav^{(t)})$, given by Eq.~(\ref{eq:inst}), to $\boldsymbol{\varsigma}_1 =(0.005~\mathrm{M}_{\odot}, 0.01~\mathrm{Gyr})$ and $\boldsymbol{\varsigma}_2=(0.01~\mathrm{M}_{\odot}, 1~\mathrm{Gyr})$. At this point, we should note that the joint PPDs for $(M,\stage)$ show strong correlations between the two parameters, as can be seen in Fig.~\ref{fig:PPD_Mt} in which we display them. Such a behaviour was expected from a theoretical point of view, but it certainly affects the efficiency of the MCMC algorithm, the instrumental laws used to sample the PPDs being symmetrical. This effect is difficult to correct manually, since the correlation coefficient appears to change from one combination $\Xv$ to the other and thus cannot be known \emph{a priori}.  

Empirically, we notice that the composite structure of the instrumental law helps us to converge if compared to simple ones, either Gaussian or uniform. This can indeed be seen from the acceptance rates for the different components of $q(\thetav^{\ast}|\thetav^{(t)})$ from one run to the other. Whereas it is always low for the uniform distribution, it varies significantly between runs for the large and the medium Gaussian. Therefore, the possibility to switch from one law to the other compensates the fact that it was not adjusted from simulation to simulation. The probabilities for law selection in Eq.~(\ref{eq:inst}) were set to $p_1=p_2=2p_3=0.4$. Of course, this is only one possible solution for the choice of the instrumental law and it might not be the optimal one. We have merely selected it for its convenience of implementation and handling.

\subsubsection{Estimates of the parameters}

The results for our MCMC simulations are reported in Table~\ref{table:simu2D}. We considered several possible combinations for $\Xv$ and give them in the second column. The third column shows the prior we used on the mass. We test the effect of uniform, Eq.~(\ref{eq:uprior}), and Gaussian, Eq.~(\ref{eq:mprior}), priors on the mass. Based on the discussion from Sect.~\ref{sect:mprior}, we shall insist on the fact that using the latter is equivalent to make a strong assumption on the value of the mass compared to the non-informative case involving a uniform distribution. We always considered a uniform prior on the age defined on the 2 -- 9 Gyr domain\footnote{With the exception of \run{1} in Table~\ref{table:simu2D}, which required a 1~Gyr lower limit for the uniform prior.}.

\begin{figure*}
\center
\includegraphics[width=\columnwidth]{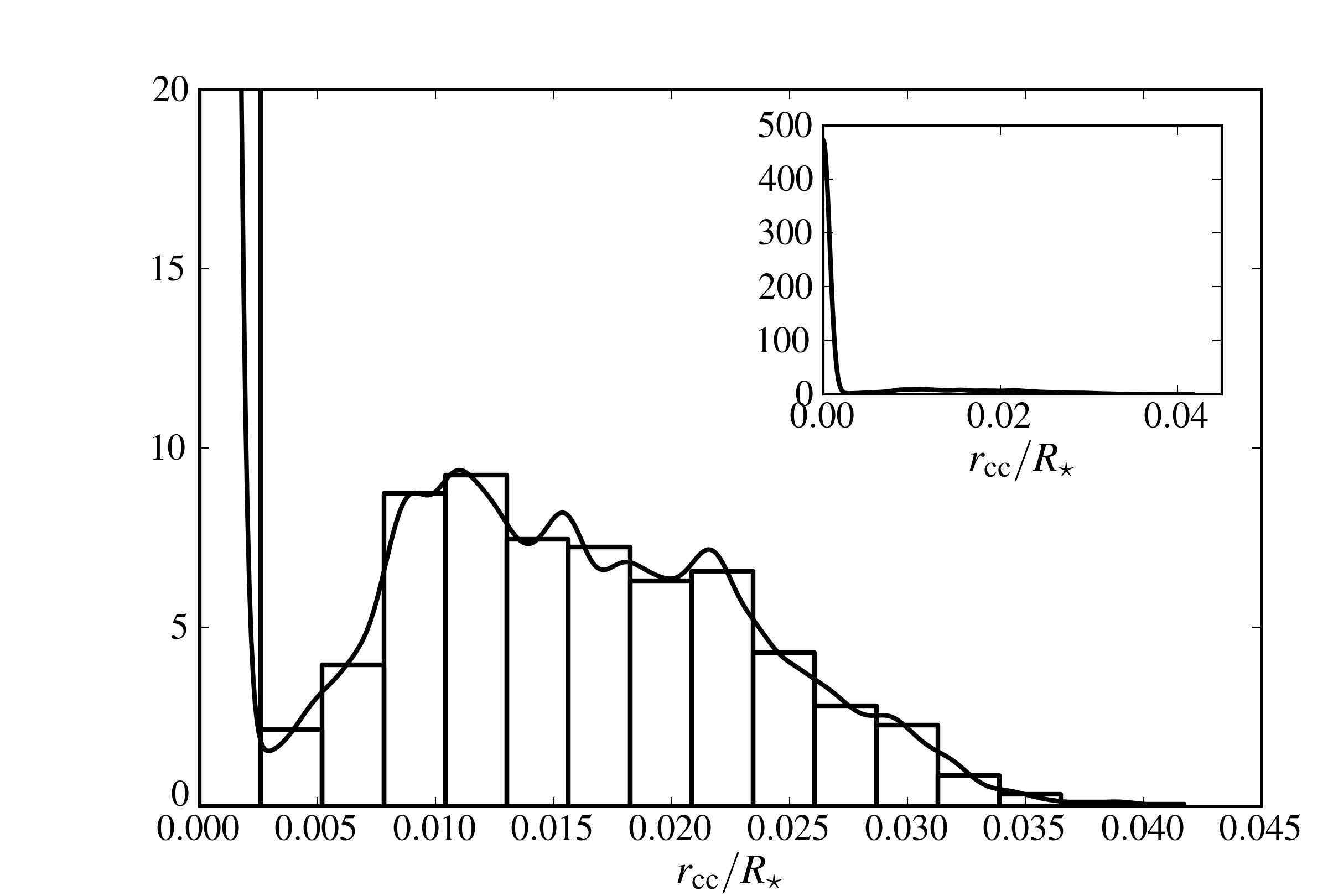}
\includegraphics[width=\columnwidth]{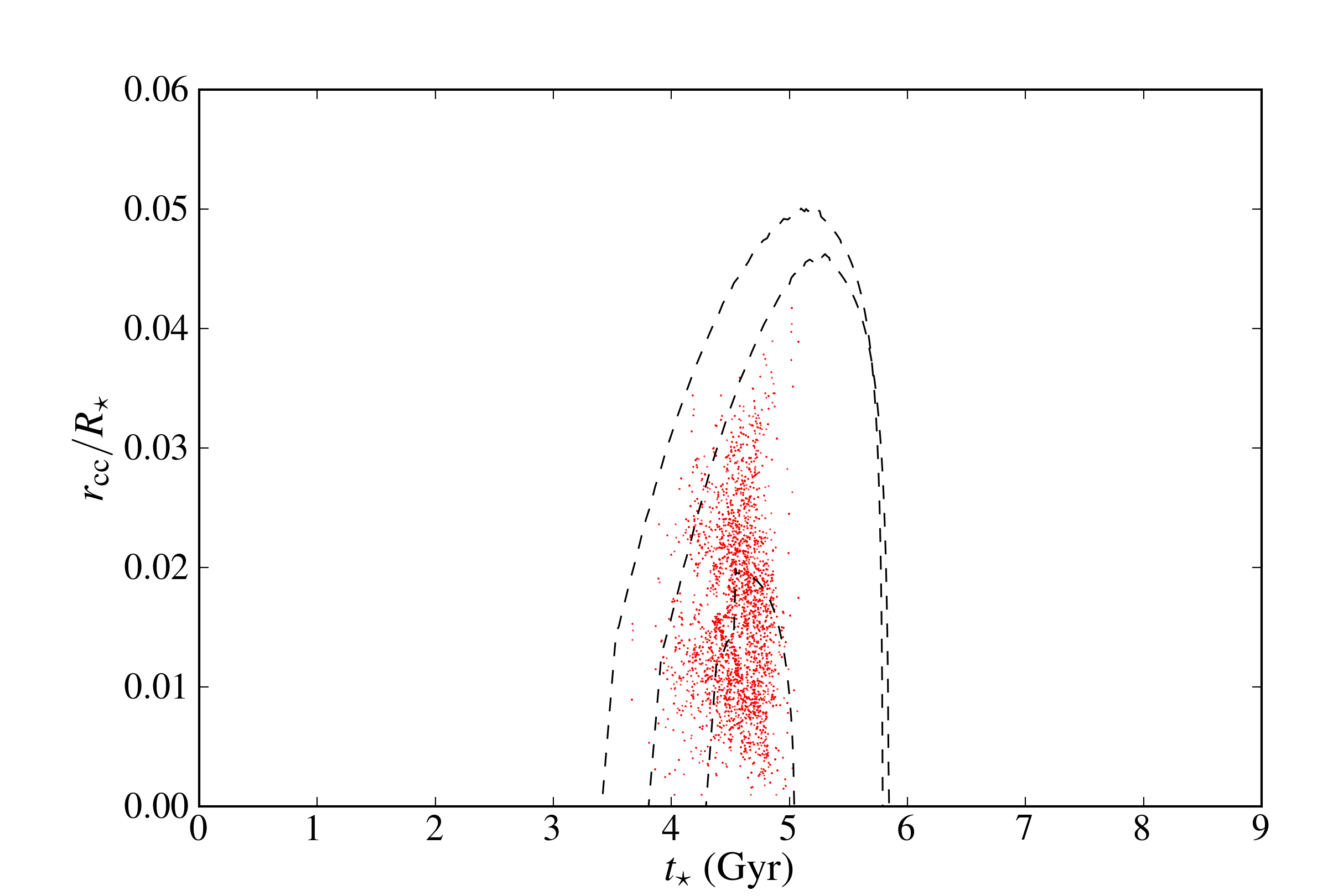}
\caption{\emph{Left panel}: Posterior probability density for the radius of the convective core of {\acena} estimated using $\Xv = (\teff,L,\ass)$ and a uniform prior on the mass (\run{5}). The PPD is represented in the form of an histogram and as a kernel-estimated continuous density. The same plot is represented with a different ordinate scale in the inset. \emph{Right panel}: Radius of the convective core as a function of the stellar age. The dots are the values obtained for {\acena} from the same MCMC simulation. The dashed lines represent the theoretical values for stars of mass, from bottom to top, $M=1.14$~{\msol}, $1.15$~{\msol} and $1.16$~{\msol}.}
\label{fig:cc2D}
\end{figure*}

From the MCMC simulation we directly obtain the joint PPDs for the parameters. This is done by simply representing the $M^{(t)}$ and $\stage^{(t)}$, $t=1,\dots,T$ in the form of a 2-dimensional histogram (see Fig.~\ref{fig:PPD_Mt}). It is also possible to obtain the marginal densities for each parameter. These are usually computed by integrating the PPD over the other parameters (sometimes called nuisance parameters). For instance, the marginal PPD for the mass is given by 

\begin{equation}\label{eq:marginal}
\pi(M|\Xv) = \displaystyle \int \pi(M,\stage|\Xv)d\stage
\end{equation}
 This integral can be approximated in the same way we did for the joint PPD. A quick inspection shows that they are all close to Gaussian distributions (which can be checked straightforwardly by computing the skewness and the kurtosis of the distribution). We can use the PM estimate and compute the Posterior Standard Deviation (PSD, square root of the Posterior Variance, noted $\sigma$) as the associated credible intervals. In this case, the PSD has the standard interpretation that a realization of the considered variable will deviate from the mean of the distribution by $\pm\sigma$ with probability $\sim$68\%. These values are given in Table~\ref{table:simu2D}, in column 4 for the mass and 5 for the age. In the context of stellar physics, these statistical summaries are perhaps the most useful products of a Bayesian analysis. The derivation of such error bars allow us to quantify precisely how the various observables and their associated uncertainties impact the stellar parameters. In our ten runs, we explored the impact of various observables on the final estimates. We always included the effective temperature and the luminosity. We then combined them with other constraints. 

The first important results that stands out is the effect of including a Gaussian prior in Eq.~\ref{eq:bayes}. In this case, the estimates of the mass are systematically lower than when a uniform prior is used. Furthermore, they are always extremely close to the value $M_{\mathrm{obs}}$, which reflects the impact of this prior. This of course leaves open a classical question that arises in Bayesian studies: is the used prior to be trusted? It is critical since including (\ref{eq:mprior}) definitely has a strong influence on the final estimates. Given the general mass-age anticorrelation, the estimated ages are also higher when the prior is Gaussian.

This picture is different when the prior on the mass is uniform and the effects of the other observed quantities appear much more clearly. To understand this, we shall first briefly examine the effect of the various combination we used for $\Xv$. The three others observational constraints we studied in detail are the radius, $R$, the average large separation, $\als$, the average small separation, $\ass$. They all lead to more precise estimates with respect to \run{1}. This effect is more important for the seismic observable than for the radius. However, it is less than when a Gaussian prior on the mass is used. 

It is clear from Table~\ref{table:simu2D} that two groups emerge among these quantities: the small separation on one hand and the radius and large separation on the other. The former, when included, leads to sensibly different estimates of the stellar parameters than the two latter. Using $\ass$ we find the highest mass for all our runs. Conversely, the estimated age is the lowest.  This seismic indicator is supposed to be mostly sensitive to the central regions of the star, which concentrate the essential of the total mass and is the locus of the nuclear burning of hydrogen, hence the best marker for stellar age. Based on these quantities, we obtain a general picture for {\acena} of a star with a mass $\sim$1.13~{\msol} and an age roughly in the range $\sim$5 -- 5.3~Gyr, of course this is only true for the physics described in Sect.~\ref{sect:model} and is susceptible of change for different prescriptions.

Both the radius and the average large separation imply lower estimates of the mass and higher estimates for the age (with respect to the estimates obtained including the average small separation in $\Xv$). For some cases, the estimates vary significantly, for instance between runs \#5 and \#6 or runs \#7 and \#8, and do not agree within their $1\sigma$ error bars (although all runs agree if we consider $3\sigma$ error bars). This discrepancy is likely to be caused, as we shall see in Sect.~\ref{sect:simu5D}, by the assumptions made on the physics of {\acena}, including in this case the setting of approximate values for $X_0$ and $\alpha$. Another source of disagreement, as discussed in Sect.~\ref{sect:obs}, is that $\als$ and $R$ are both affected by the surface layers of the star and reproducing them would imply the use of a stellar evolution code that models accurately these regions. Finally, we also note that the effect of $\als$ is the same as $R$ on the PM estimates, although lesser in magnitude than including a Gaussian prior on the mass, whose observational determination might  also suffer from a bias related to the poorly known stellar surface (Sect.~\ref{sect:mprior}).

\subsubsection{Structure of {\acena}}\label{sect:2D_struct}

\begin{figure*}
\center
\includegraphics[width=\columnwidth]{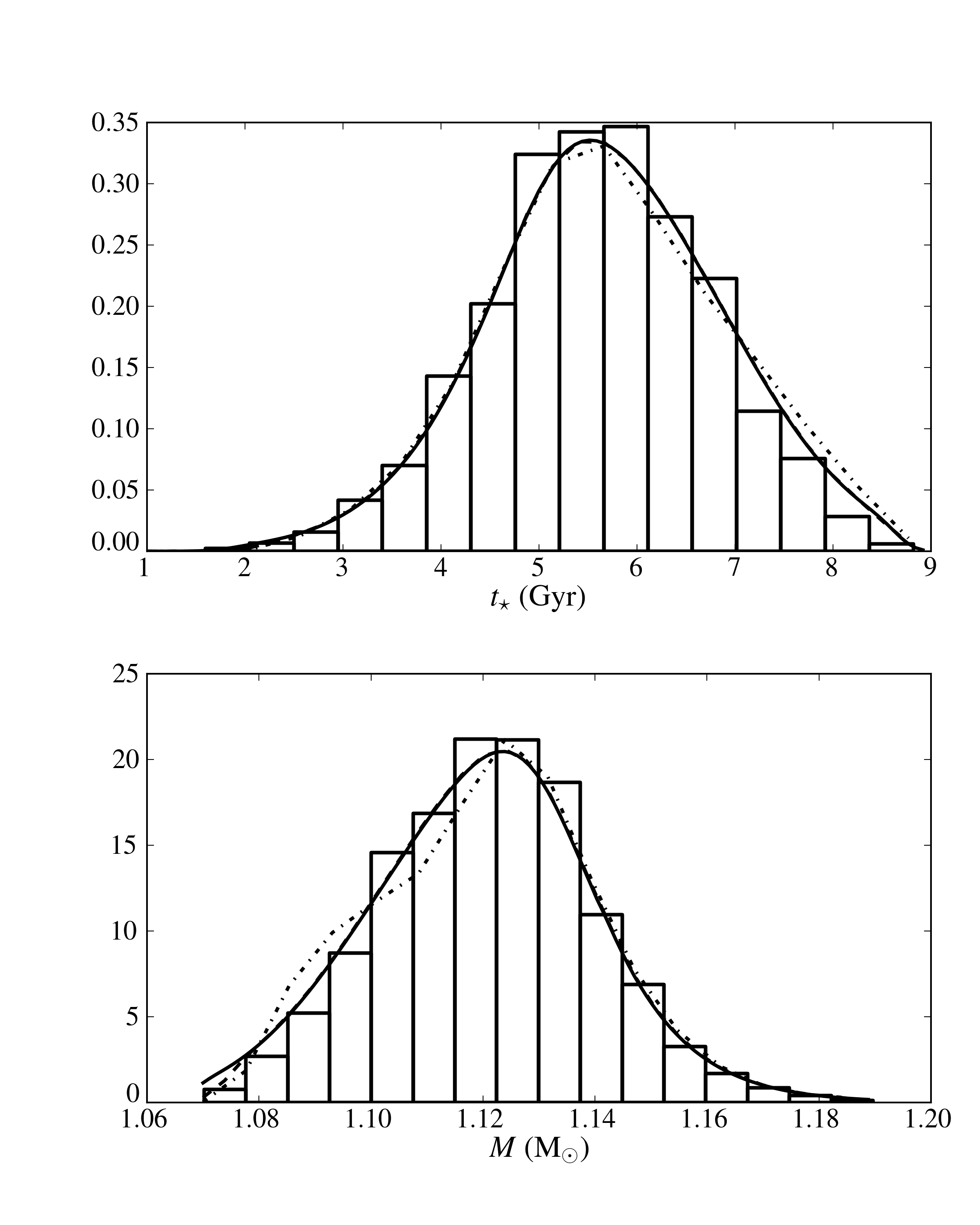}
\includegraphics[width=\columnwidth]{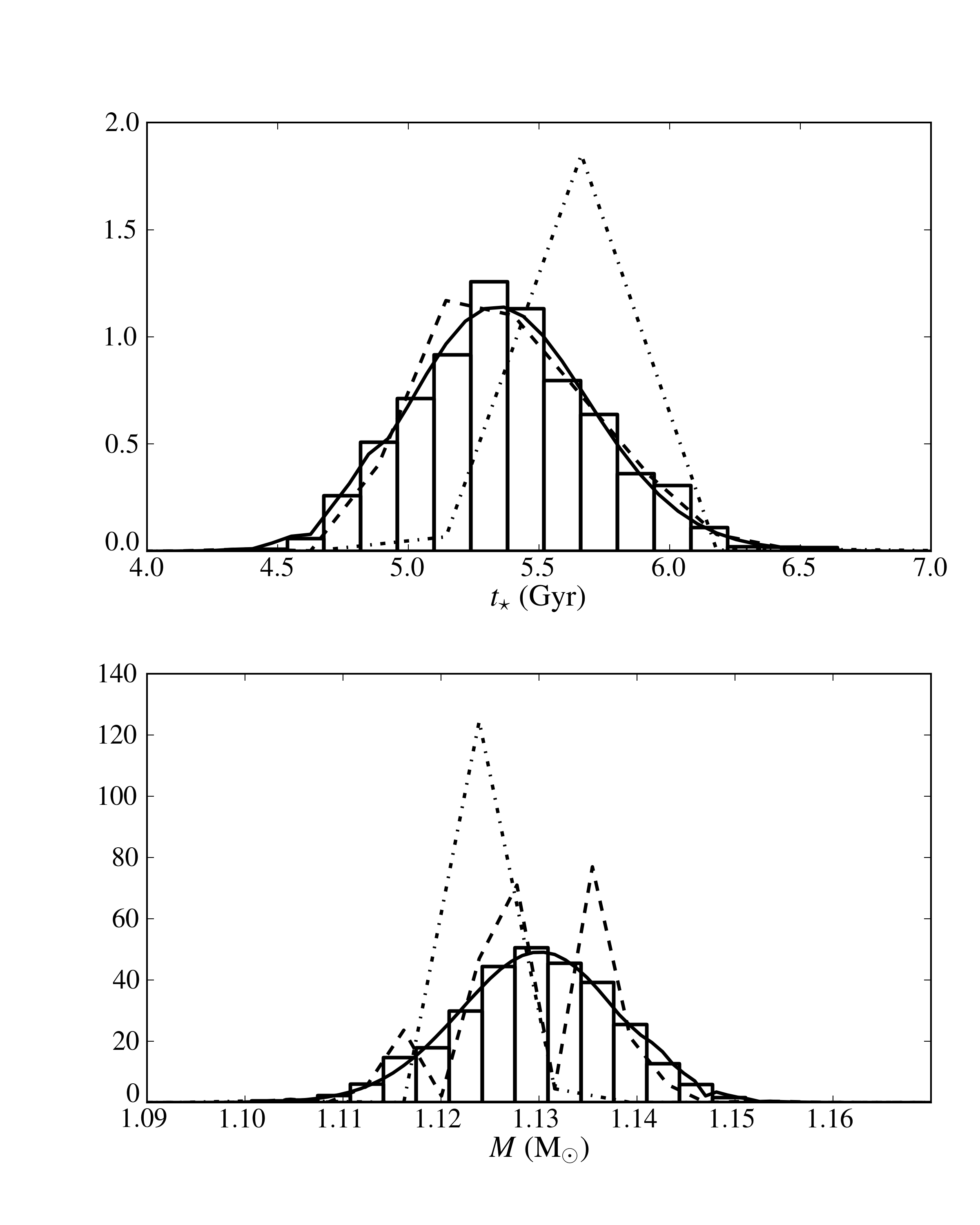}
\caption{\emph{Left column}: marginal PPDs for the mass (bottom graph) and the age (top graph) of {\acena}. The observational constraints are $\Xv = (\teff,L)$ and the prior on the mass is uniform. The histogram represent the results from the MCMC simulation. The lines represent the same marginal PPD computed from grids \#2 (full line), \#3 (dashed line) and \#4 (dashed-dotted line). \emph{Right column}: same as left column with $\Xv = (\teff,L,\ass)$.}
\label{fig:sgrid2D}
\end{figure*}

 These variations on the estimated mass and age with respect to $\Xv$ affect the general picture we can draw of the internal structure of {\acena}. A good example concerns the search for a convective core that has been a long-standing problem for this star \citep{Turcotte98,Guenther00,Morel00,Thevenin02,Thoul03,Eggenberger04,Miglio05,Montalban06}. It is possible to obtain as an output of an MCMC simulation the PPD for any function of the parameters\footnote{Once we have estimated the joint PPD for $\thetav$, we also have an estimate of the PPD of any quantity $h$ such that $h=h(\thetav)$.}. We can thus estimate the probability of existence of a convective core $P(r_{cc} > 0)$ (or similarly $P(m_{cc} > 0)$), with $r_{cc}$ the radius of the convective core (and $m_{cc}$ its mass). The distribution for the radius of the convective core obtained from \run{5} is shown as an example in Fig.~\ref{fig:cc2D}. We plotted in the form of an histogram and as a continuous density obtained from kernel estimation. However, we note that it is delicate to find an adequate representation since it is obviously discontinuous at $r_{cc} = 0$.

In the runs not including seismic constraints, we find a probability of having a convective core of 10\% and 2.5\% for \#1 and \#2 respectively. In \run{3}, for which the Gaussian prior on the mass was used, this probability becomes null. We shall note right now that, regardless the observation vector chosen, we never identify a model with a convective core when using this prior. For a uniform prior on the mass, we can again draw a line between estimates, depending whether $\ass$ or $\als$ has been included in $\Xv$. In the former case, runs \#5 and \#7, we find a probability for the existence of a convective core of 16.5\% and 11.2\% respectively. In the latter case, it becomes null. As we shall see in Sect~\ref{sect:5D_struct} this is partly due to the sensitivity of $\ass$ to the innermost regions of the star, but it is also a bias due to our assumptions on $Z_0$ and $X_0$. 

These results also allow us to derive an upper limit on the radius and the mass of a possible convective core. If we base ourselves on \run{5}, we find $r_{cc} \lesssim 0.040$~R$_{\star}$ and $m_{cc} \lesssim 0.015$~M$_{\star}$ (these values are compatible with the results from \run{7}). Note also that the distribution of $r_{cc}$ is not Gaussian anymore. We can see here an obvious advantage of the Bayesian approach: if one wants to estimate uncertainties on such physical parameters, it is not possible to rely on point-estimation method such as likelihood maximum and use, for instance, Hessian inversion to derive them. One has to sample the entire distribution for this quantity to be in position to provide with robust statistics. 

From Fig.~\ref{fig:cc2D}, we can also obtain a general picture of the possible evolutionary state of the convective core of {\acena}. It displays the size of the convective core as a function of time of the stellar age, the dots are the results from the MCMC simulation and the dashed line gives $r_{\mathrm{cc}} = r_{\mathrm{cc}}(\stage)$ for $M=1.14$~{\msol}, $1.15$~{\msol} and $1.16$~{\msol}. It appears that the convective cores are in a growing phase. This mirror the fact that we tend to find younger age estimate when using $\ass$. The growing phase is relatively longer than the receding one, explaining the significant probability obtained for the convective core existence. Similarly, the lower percentage for the existence of a convective core in \run{7} is explained by the higher age and only a handful of models are being selected which are in the shorter time-scale receding phase of the convective core. 

\subsubsection{Comparison with grid-based integration}\label{sect:grids2D}

Sampling only $M$ and $\stage$ allowed us to compute a very dense grid in the region of interest of the space of parameters, Using such a grid, it is straightforward to integrate directly the denominator in the right-hand side of Eq.~(\ref{eq:bayes}), which simply is
\begin{equation}
\displaystyle m(\Xv) = \int_{\pspace} \pi(\thetav)\mathscr{L}(\Xv|\thetav) d\thetav,
\end{equation}
using classical quadrature formula (the trapezoidal rule in this case). Similarly, we can integrate out the nuisance parameters to obtain the marginal PPDs using formula of the same type as (\ref{eq:marginal}).

To proceed to this comparison, we computed a grid of approximate size $N_g = 216\times 216$ in the $(M - \stage)$ plane. It was chosen so that the total number of computed models is close to 45000, which is the size of our MCMC sample when the typical burn-in sequence of 5000 models is removed. In fact, the total number of computed models is lower than $N_g$ because not all points in the $(M - \stage)$ plane correspond to stable solutions for our stellar evolutionary code. Stars massive enough might indeed reach the giant branch well before the required $\stage^{\ast}$. Models producing numerical convergence errors during these phases of the stellar evolution were discarded. The effective number of computed models in our grid is $N_g^{\mathrm{eff}} = 45533$.

\input{Tables/simu5D}

Figure~\ref{fig:sgrid2D} shows a comparison between the marginal PPDs obtained with our MCMC simulations and from direct integration from this grid. We selected two example cases with $\Xv = (\teff,L)$ and $\Xv = (\teff,L,\ass)$. The MCMC and grid-integrated distributions are almost identical. This can be checked graphically or by comparing the first moments of the marginal densities given in Table~\ref{table:simu2D}. They agree within a few percents at most. This was expected and is a good test of the self-consistency of our Bayesian approach that two radically different numerical strategies converge. At such a level of agreement, the choice of the method should be made on the basis of the computational performances. Producing dense grids of stellar models in 2-dimensional parameter spaces is relatively inexpensive, therefore this should be the approach of choice compared to the MCMC.

The fact that we are using a dense grid is critical. When considering 5-dimensional parameter spaces we will not have this possibility. Therefore, we use the computational flexibility we have in a lower-dimensional parameter space to illustrate the effect of sparser sampling on the calculation of the PPD. For this, we have selected subgrids in $\pspace$ by removing points from our original one. They have dimensions $216\times216$, $108\times108$, $32\times32$ and $16\times16$. As mentioned before, some models could not be computed with ASTEC and the effective numbers of points are, respectively, 11\,397, 936 and 245. We repeated the integration of the PPDs to obtain the corresponding marginal densities and their first two moments . We note that the original grid covers the $(M-\stage)$ plane with regular cells of dimension $5.5\times 10^{-4} M_{\odot}\times 37~\mathrm{Myr}$. The sparsest subgrid has cells of size $7.7\times 10^{-3} M_{\odot}\times 518~\mathrm{Myr}$. Even this large-sampling units offer very acceptable coverage of the space of parameters, in particular in regard with commonly used grids which use steps in mass\footnote{The problem of the stellar age sampling is a bit particular. Most of the time, these grids are database of \emph{stellar evolutionary tracks}. This means that the stellar age is imposed by the stellar evolution code itself and not the user. We still needed to maintain a strict control of this parameter in order to compare properly with MCMC algorithm.} $\gtrsim 0.01$~{\msol} \citep[see e.g.][]{Quirion10}.

The main result is that downsizing the sampling can significantly affect the outcome of the estimation process. The values of the posterior means and standard deviations for all selected subgrids are reported in Table~\ref{table:simu2D}. For the first two subgrids, the results agree fairly well with the MCMC simulations and the dense-grid integration. However, the results for the sparsest grid show significant discrepancies with these reference estimates. The estimated posterior means can depart from the reference ones by a few percent (around 3-5\% for the worst cases). We thus have a numerically-induced loss of accuracy. This down-sampling also affects the estimates of the uncertainties (which can be seen as the precision on the parameters and should be distinguished from the accuracy). In some cases, they can be overestimated. For instance, in \run{1} the uncertainty on the age is 30\% larger when using grid \#4. Conversely some uncertainties can also be underestimated. This is significant since we see departures ranging from the MCMC and dense-grid estimates ranging from a factor $\sim$3 (\run{5}) to several orders of magnitudes (runs \#6, \#8 or \#10). These very low uncertainties sometimes obtained with grid \#4 simply means that only few points in our sampling scheme correspond to a non-zero values of the PPD. In these cases, the numerical integral we perform to compute the second moment returns unrealistic low values. This can be seen as a numerical error introduced by the sampling itself.

In general, and unsurprisingly, the sharper the peaks of the densities, the more important this phenomenon is. This reflects that the sampling in the parameter space is not dense enough to capture properly the variations of the PPD. This in turn impacts the numerical approximation to the integrals as can be seen in Fig.~\ref{fig:sgrid2D}. It appears clearly that the estimated densities depart strongly from the correctly-sampled cases (MCMC or dense grids) for $\Xv = (\teff,L,\ass)$, less so for $\Xv = (\teff,L)$.

Note that we chose the smallest subgrid because it has $N_g^{\mathrm{eff}}\sim16^2=256$ points. This is representative of the density of points we have in our 5-dimension grids. It is also noteworthy that this behaviour is only indicative of the problem we might encounter when dealing with higher dimension. It is not possible to extrapolate fully the results of the two-dimensional to the five-dimensional case.

%% file: Tables/simu2D_all.tex
\begin{table*}
\begin{center}
\caption{Inferred stellar parameters in the $(M - \stage)$ plane of the parameter space constrained using various combination, $\Xv$, of the observables. First column: the number of the run, second column: observational constraints included in $\Xv$, third column: prior used on the mass. The next columns give the estimates for these quantites in the form of the MP estimate and the associated credible $1\sigma$ credible interval. Columns 4 and 5: estimates obtained with our MCMC algorithm, columns 6 and 7: estimates obtained by computing the PPD on ($216\times216$) grid sampling regularly the $(M - \stage)$ plane, columns 8 to 13: same with ($108\times108$), ($32\times32$) and ($16\times16$) subgrids in the $(M - \stage)$ plane. Note that the dimension of the grid is indicative, the effective number of points used is lower since some models could not be computed (see text).}
\label{table:simu2D}
\begin{tabular}{lcccccccccc}
\toprule %\hline

      &     &        &\multicolumn{2}{c}{MCMC} & \multicolumn{2}{c}{Grid \#1 ($216\times216$)}& \multicolumn{2}{c}{Grid \#2 ($108\times108$)} \\
\cmidrule(r){4-5}  \cmidrule(r){6-7} \cmidrule(r){8-9}  
Run \#&$\Xv$&$\pi(M)$& $\hat{M}$ ({\msol})& $\hat{\stage}$ (Gyr)& $\hat{M}$ ({\msol})& $\hat{\stage}$ (Gyr)& $\hat{M}$ ({\msol})& $\hat{\stage}$ (Gyr)\\

\midrule
1 &$(L,\teff)                                                  $&Uniform &$1.121 \pm 0.019$&$5.63 \pm 1.14$&$1.120 \pm 0.020$&$5.67 \pm 1.20$&$1.120 \pm 0.020$&$5.68 \pm 1.20$\\
2 &$(L,\teff,R)                                                $&Uniform &$1.116 \pm 0.011$&$5.98 \pm 0.52$&$1.116 \pm 0.012$&$5.99 \pm 0.53$&$1.116 \pm 0.012$&$5.99 \pm 0.53$\\
3 &$(L,\teff)                                                  $&Gaussian&$1.107 \pm 0.007$&$6.47 \pm 0.44$&$1.107 \pm 0.007$&$6.48 \pm 0.44$&$1.107 \pm 0.007$&$6.48 \pm 0.44$\\
4 &$(L,\teff,R)                                                $&Gaussian&$1.108 \pm 0.006$&$6.34 \pm 0.28$&$1.108 \pm 0.006$&$6.34 \pm 0.28$&$1.108 \pm 0.006$&$6.34 \pm 0.28$\\
5 &$(L,\teff,\langle\delta\nu\rangle)                          $&Uniform &$1.130 \pm 0.008$&$5.03 \pm 0.39$&$1.131 \pm 0.008$&$5.00 \pm 0.39$&$1.130 \pm 0.008$&$5.03 \pm 0.39$\\
6 &$(L,\teff,\langle\Delta\nu\rangle)                          $&Uniform &$1.109 \pm 0.009$&$6.35 \pm 0.38$&$1.111 \pm 0.009$&$6.28 \pm 0.37$&$1.111 \pm 0.009$&$6.28 \pm 0.37$\\
7 &$(L,\teff,R,\langle\delta\nu\rangle)                        $&Uniform &$1.130 \pm 0.008$&$5.36 \pm 0.34$&$1.130 \pm 0.008$&$5.35 \pm 0.35$&$1.130 \pm 0.008$&$5.36 \pm 0.35$\\
8 &$(L,\teff,R,\langle\Delta\nu\rangle)                        $&Uniform &$1.103 \pm 0.007$&$6.62 \pm 0.29$&$1.103 \pm 0.007$&$6.60 \pm 0.27$&$1.103 \pm 0.007$&$6.60 \pm 0.27$\\
%9 &$(L,\teff,R,\langle\delta\nu\rangle,\langle\Delta\nu\rangle)$&Uniform &$\pmb{1.102 \pm 0.006}$          &$\pmb{6.63 \pm 0.26}$       &$1.111 \pm 0.006$         &$6.27 \pm 0.25$       &$1.111 \pm 0.006$&$6.27 \pm 0.25$\\
9&$(L,\teff,R,\langle\delta\nu\rangle)                         $&Gaussian&$1.116 \pm 0.006$&$5.98 \pm 0.25$&$1.115 \pm 0.006$&$5.99 \pm 0.25$&$1.115 \pm 0.006$&$6.00 \pm 0.25$\\
10&$(L,\teff,R,\langle\Delta\nu\rangle)                        $&Gaussian&$1.103 \pm 0.005$&$6.58 \pm 0.20$&$1.104 \pm 0.005$&$6.56 \pm 0.20$&$1.104 \pm 0.005$&$6.56 \pm 0.20$\\
%12&$(L,\teff,R,\langle\delta\nu\rangle,\langle\Delta\nu\rangle)$&Gaussian&$\pmb{1.103 \pm 0.005}$          &$\pmb{6.58 \pm 0.20}$       &$\pmb{1.108 \pm 0.005}$          &$\pmb{6.38 \pm 0.19}$       &$1.108 \pm 0.005$&$6.38 \pm 0.19$\\

\bottomrule
\end{tabular}
\vspace{.2cm}
\begin{tabular}{lcccccc}
\toprule %\hline

      &     &        &\multicolumn{2}{c}{Grid \#3 ($32\times32$)} & \multicolumn{2}{c}{Grid \#4 ($16\times16$)} \\
\cmidrule(r){4-5}  \cmidrule(r){6-7} 
Run \#&$\Xv$&$\pi(M)$& $\hat{M}$ ({\msol})& $\hat{\stage}$ (Gyr)& $\hat{M}$ ({\msol})& $\hat{\stage}$ (Gyr)\\

\midrule
1 &$(L,\teff)                                                  $&Uniform &$1.120 \pm 0.020$&$5.68 \pm 1.19$&$1.120 \pm 0.020$&$5.70 \pm 1.21  $\\
2 &$(L,\teff,R)                                                $&Uniform &$1.116 \pm 0.012$&$5.99 \pm 0.53$&$1.115 \pm 0.012$&$6.04 \pm 0.55  $\\
3 &$(L,\teff)                                                  $&Gaussian&$1.107 \pm 0.007$&$6.48 \pm 0.44$&$1.106 \pm 0.007$&$6.45 \pm 0.44  $\\
4 &$(L,\teff,R)                                                $&Gaussian&$1.108 \pm 0.006$&$6.35 \pm 0.29$&$1.102 \pm 0.005$&$6.64 \pm 0.23  $\\
5 &$(L,\teff,\langle\delta\nu\rangle)                          $&Uniform &$1.131 \pm 0.008$&$5.03 \pm 0.39$&$1.131 \pm 0.008$&$5.02 \pm 0.39  $\\
6 &$(L,\teff,\langle\Delta\nu\rangle)                          $&Uniform &$1.113 \pm 0.009$&$6.19 \pm 0.37$&$1.101 \pm 3\times10^{-5}$&$6.70 \pm 1\times10^{-3} $\\
7 &$(L,\teff,R,\langle\delta\nu\rangle)                        $&Uniform &$1.129 \pm 0.008$&$5.38 \pm 0.33$&$1.124 \pm 0.002$&$5.65 \pm 0.12  $\\
8 &$(L,\teff,R,\langle\Delta\nu\rangle)                        $&Uniform &$1.102 \pm 0.004$&$6.66 \pm 0.18$&$1.101 \pm 2\times10^{-5}$&$6.70 \pm 9\times10^{-4} $\\
9&$(L,\teff,R,\langle\delta\nu\rangle)                         $&Gaussian&$1.115 \pm 0.005$&$5.99 \pm 0.24$&$1.120 \pm 0.008$&$5.81 \pm 0.34  $\\
10&$(L,\teff,R,\langle\Delta\nu\rangle)                        $&Gaussian&$1.101 \pm 0.002$&$6.69 \pm 0.07$&$1.101 \pm 5\times10^{-7}$&$6.70 \pm 2\times10^{-5} $\\
\bottomrule
\end{tabular}
\end{center}
\end{table*}

%% file: Figures/PPD_Mt.tex
\begin{figure*}
\begin{center}
\includegraphics[width=\textwidth]{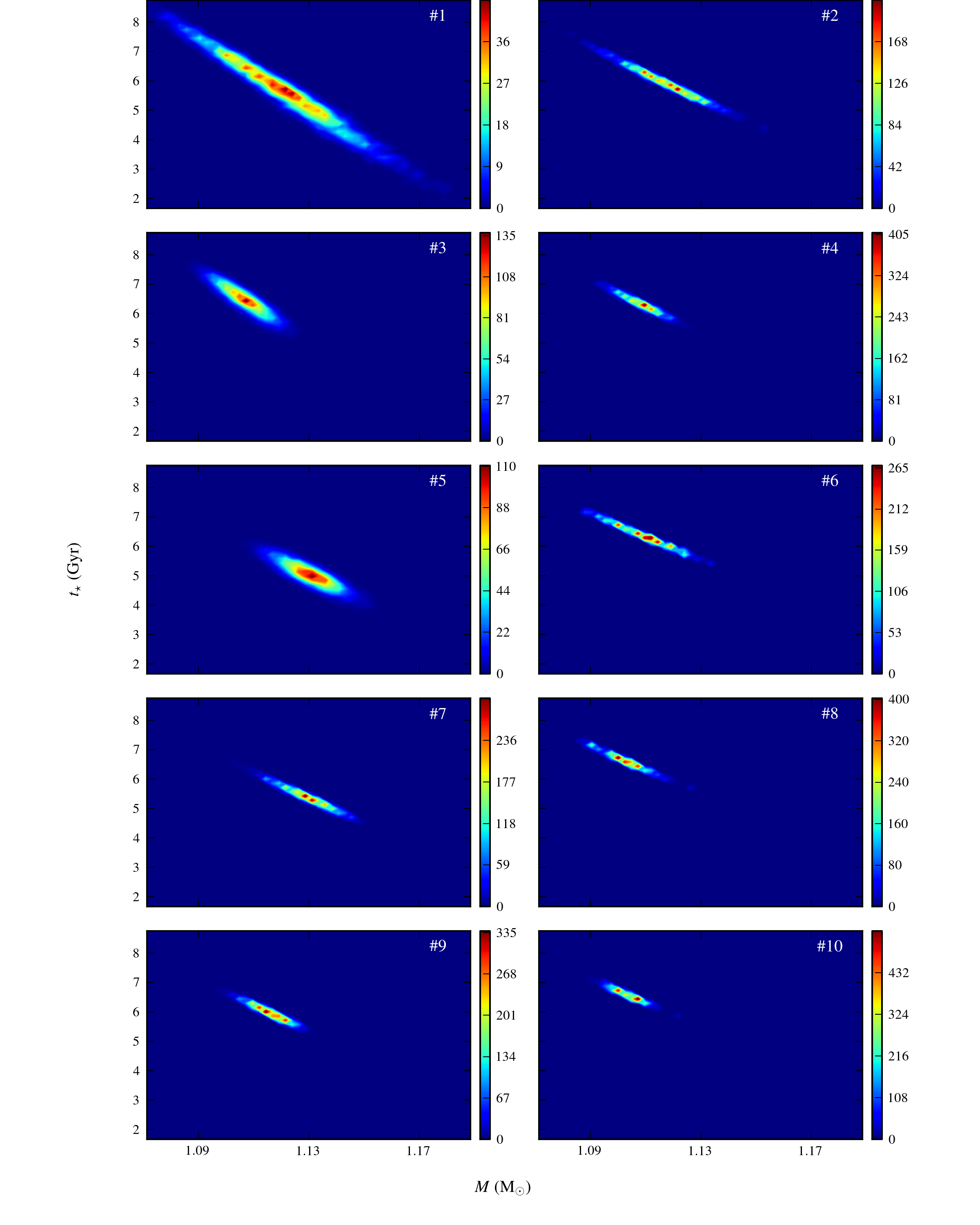}
\caption{Joint Posterior Probability Densities for the couple $(M,\stage)$ obtained from our MCMC simulations. Each one is labeled with the number given in the first column of Table~\ref{table:simu2D}.}
\label{fig:PPD_Mt}
\end{center}
\end{figure*}

%% file: Tables/simu5D.tex
\begin{table*}
\begin{center}
\caption{Inferred stellar parameters constrained using seismic and non-seismic observables. First column: the number of the run, second column: observational constraints included in $\Xv$, third column: prior on the mass, colums 4-8: PM estimates and associated credible intervals (the distributions are assumed Gaussian) for $M$, $\stage$, $Z_0$, $\alpha$ and $X_0$ respectively.}
\label{table:simu5D}
\begin{tabular}{lcccccccc}
\toprule %\hline

Run \#&$\Xv$&$\pi(M)$& $\hat{M}$ ({\msol})& $\hat{\stage}$ (Gyr) & $Z_0$ & $\alpha$ & $X_0$\\

\midrule

12&$[L, \teff, R, \langle \delta \nu \rangle]$                            &Uniform  &$1.12\pm 0.03$ &$4.8\pm0.5$ &$0.027\pm0.003$ &$1.7\pm0.1$  &$0.69\pm0.02$ \\

13&$[L, \teff, R, \langle \delta \nu \rangle]$                            &Gaussian &$1.105\pm0.007$ &$4.8\pm0.5$ &$0.026\pm0.003$  &$1.62\pm0.08$ &$0.68\pm0.02$ \\

14&$[L, \teff, R, \langle \delta \nu \rangle, \langle \Delta \nu \rangle]$&Gaussian& $1.102\pm0.005$ & $4.8\pm0.5$ & $0.026\pm0.003$ & $1.60 \pm 0.10$ & $0.69\pm0.02$ \\

\bottomrule
\end{tabular}
\end{center}
\end{table*}

%% file: Sections/Subsections/5D.tex
\subsection{MCMC simulations with $\pmb{\dim(\pspace) = 5}$}\label{sect:simu5D}

\subsubsection{Implementation details}

Our next step is to increase the number of parameters left free to vary. As noted in Sect.~\ref{sect:simu2D_implementation}, considering only the stellar mass and age imply to make strong assumptions on other parameters. In particular we needed to extrapolate the mixing-length and the initial hydrogen mass fraction from their solar values, which is always a delicate procedure that should ideally be restricted to stars lying in a close vicinity of the Sun in the space of parameters \citep[such as solar twins see e.g.,][]{Bazot11}. Most of the grids of stellar parameters offer more than two values to sample from \citep[for a recent example, see][]{Quirion10}. However, increasing the dimension of $\pspace$ comes to a computational cost, namely that, all other things held constant, the number of values to consider in a volume element grows exponentially.

Our goal here is two-fold. We first want to estimate $M$, $\stage$, $Z_0$, $\alpha$ and $X_0$ together for {\acena} and determine the impact of relaxing the last three in regard with the results from Sect.~\ref{sect:simu2D}. We then try to evaluate whether we can gain in efficiency when using the MCMC algorithm with respect to the grid approach in the 5-dimensional case.

The physical settings of ASTEC remain the same. The initial metallicity, the mixing-length parameter and the initial hydrogen mass fraction are now left to vary. The relevant domains are $[0.015,0.036]$ for $Z_0$, $[1.0,3.0]$ for $\alpha$ and $[0.6,0.8]$ for $X_0$. These limits were set after several trial runs. The Gaussian components of the instrumental law $q(\thetav^{\ast}|\thetav^{(t)})$ were set to $\pmb{\varsigma}_1 =(5\times10^{-3}~\mathrm{M}_{\odot}, 0.1~\mathrm{Gyr},5\times10^{-5},1\times10^{-2},1\times10^{-3})$ and $\pmb{\varsigma}_2=(2.5\times10^{-2}~\mathrm{M}_{\odot}, 0.5~\mathrm{Gyr},5\times10^{-4},5\times10^{-2},1\times10^{-2})$. 

In order to limit the computational costs, we restricted the combinations we considered in the observable space for $\Xv$. We used the full set of non-seismic constraints and tested them together with $\ass$ and $\als$. For each case, we considered a Gaussian prior on the mass. For $\Xv = (\teff,L,R,\langle \delta\nu \rangle)$ we also studied the effect of a uniform prior on the mass. The prior on the age, the metallicity, the mixing-length parameter and the initial hydrogen fraction are always uniform. Our Markov Chains are approximately three to four time longer than in the 2-dimensional case, ranging within 150\,000-200\,000 iterations. The burn-in sequence is also longer. After testing several values for this latter, we chose to discard a generic 10\,000-iteration sequence. We also used 8 parallel chains for each run, increasing the computational cost but providing with a possibility to check the convergence of our simulations with some robustness.

\subsubsection{Estimates of the parameters}
\begin{figure*}
\center
\includegraphics[width=\columnwidth]{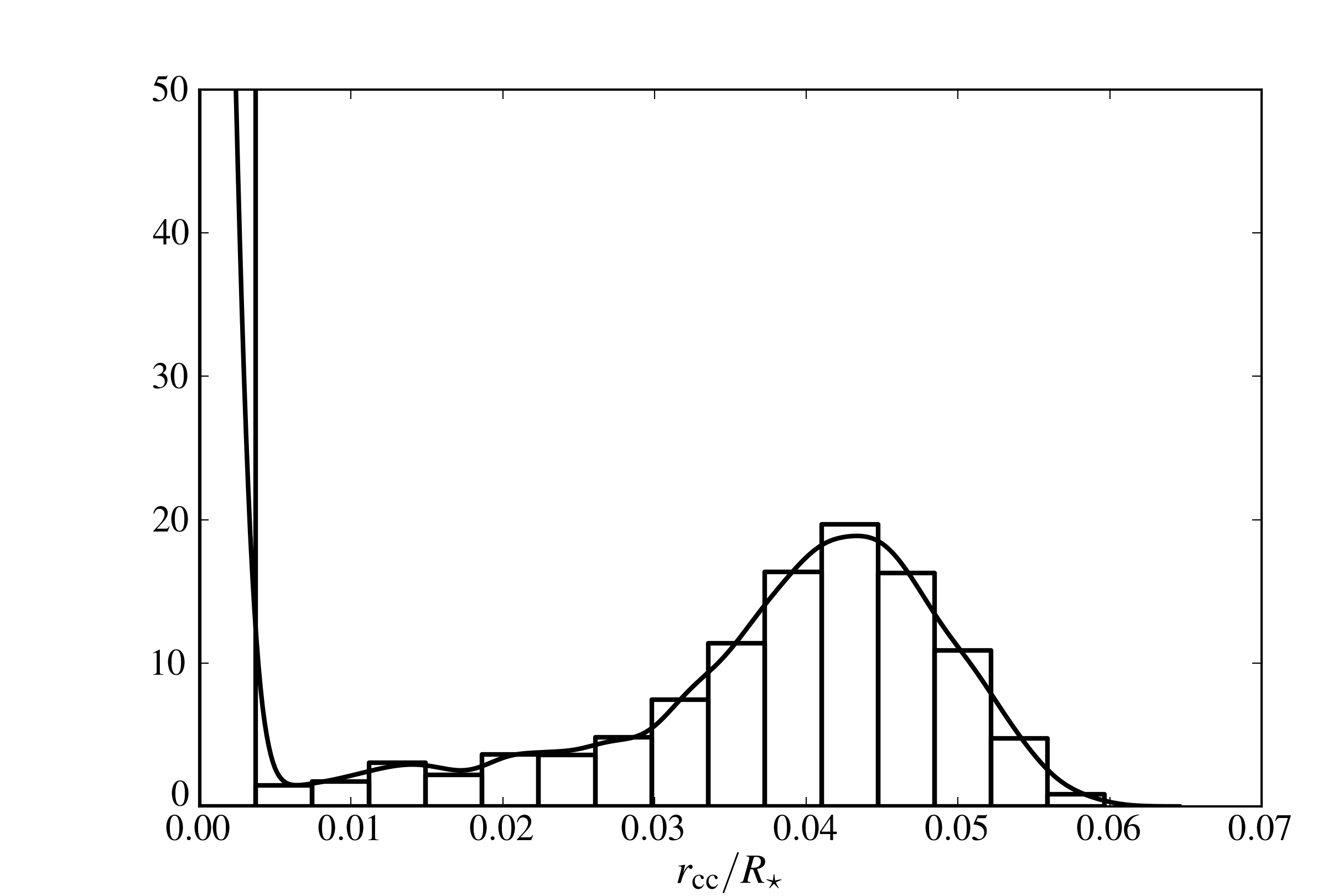}
\includegraphics[width=\columnwidth]{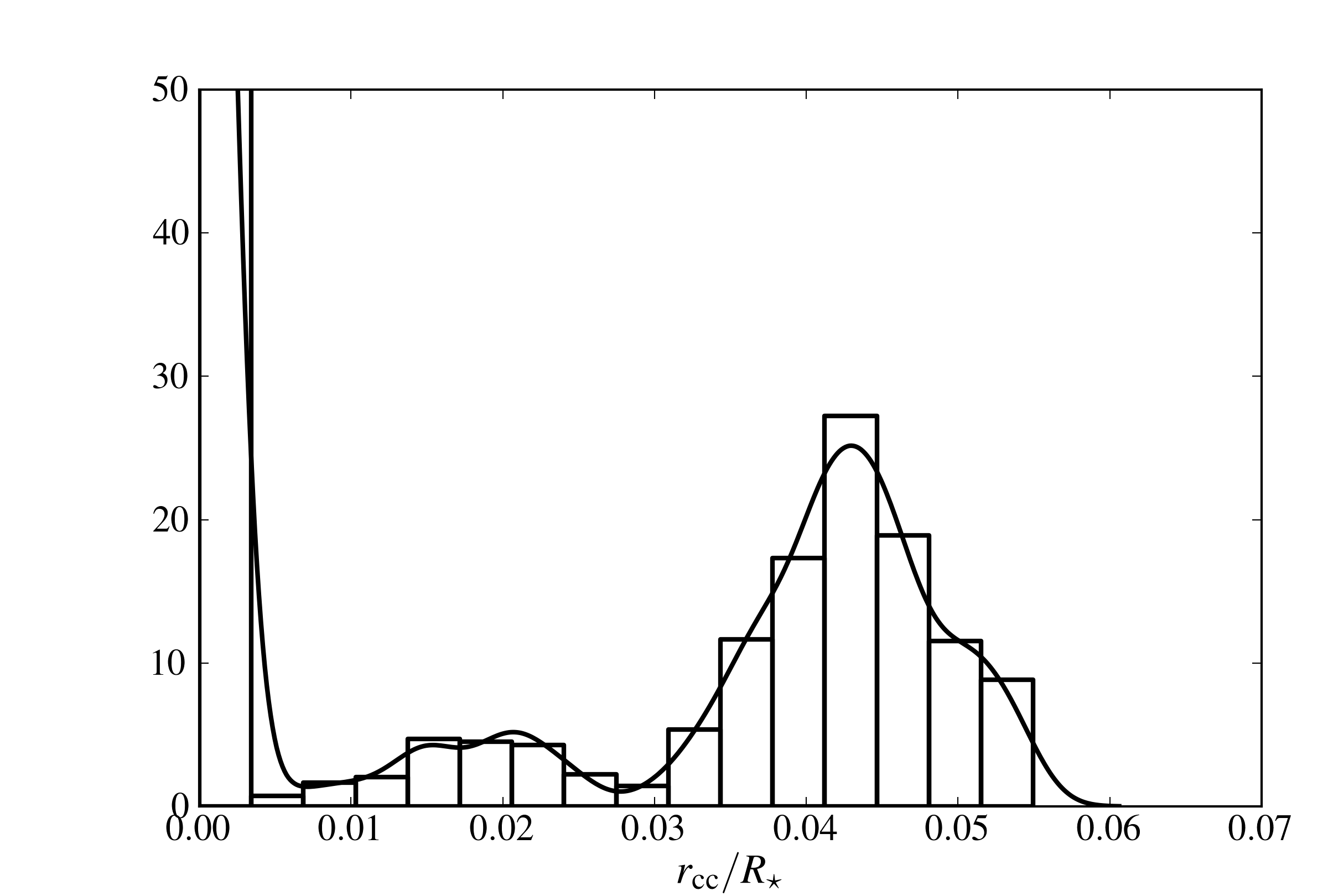}
\caption{\emph{Left panel}: Posterior probability density for the radius of the convective core of {\acena} estimated using $\Xv = (\teff,L,R,\ass)$ and a Gaussian prior on the mass (\run{12}). The PPD is represented in the form of an histogram and as a kernel-estimated continuous density. \emph{Right panel}: Same as left panel for $\Xv = (\teff,L,R,\ass,\als)$ (\run{13})}
\label{fig:cc5D}
\end{figure*}

First and foremost, it should be noted that the efficiency of all the strategies decreases when $\dim(\pspace)$ increases. This has practical incidences on the subsequent statistical analysis. It becomes difficult to approximate correctly the PPD for five stellar parameters in a reasonable amount of time. This is due to two entangled factors i) stellar models take relatively long to compute and this can become a limiting factor when increasing the number of points to sample in $\pspace$, ii) with increasing dimension, the shape of probability densities might become complex, this is expected to be even more critical for stellar models, which are highly non-linear in their parameters, hence being more difficult to sample properly from the simpler distributions used for the instrumental law.

A simple manifestation of these effects can first be seen in Fig.~\ref{fig:mPPD_5D}. We plotted the marginal densities of the five parameters for \run{13}. It is obvious by simply comparing these densities that one has to explore a much larger volume in $\pspace$. This translates into a much higher uncertainty on the mass obtained from \run{12} (compared to \run{7}) or on the age from \run{13} (compared to \run{9}). Second, the departures from gaussianity become more important than what was observed in the 2-dimensional case. Third, one has to consider that these values have been obtained for chains of length $\sim$$150\,000$. We controlled our results using multiple parallel chains and noticed that, in some cases, convergence does not seem to be reached. This might be due to one particular chain being stuck in a local maximum or not exploring efficiently enough the parameter space. Furthermore, the variance between the multiple chains is higher than in the 2-dimensional case (as could be seen using the between-chain variance $B_T$ defined in Appendix~\ref{app:conv}). Practically, this is reflected by the rounding of our estimates in Table~\ref{table:simu5D}. In more intuitive terms, this can be interpreted as a loss of numerical precision between the 2 and 5-dimensional cases, \run{12} being a clear example of this effect.

Table~\ref{table:simu5D} also offers good insight on the impact of observational data. The most striking fact is the huge effect the prior on the mass has. With a uniform prior, the precision on the mass decreases by almost an order of magnitude. We should also note the numerical advantage offered by the use of the Gaussian prior: by simplifying the target PPD  it favours greatly the convergence properties of the MCMC algorithm.

Since the mass measurements can be interpreted in terms of prior density, their impact is relatively straightforward to evaluate. The relation between the other observables and the stellar parameters is much more difficult to unveil. Theoretically, it is expected that the average small separation should be a good constraint for the stellar age. To test this hypothesis without performing as many runs as we did in Sect.~\ref{sect:simu2D} (and also because the use of $\als$ alone was impractical, see below), we performed another MCMC run (\run{14}) with the observable vector $\Xv = (\teff,L,R,\ass,\als)$. The results comparison with \run{13} is particularly interesting. It appears clearly that the impact of the small separation is strong, and in fact much more constraining than the large separation. Comparing \run{12} an \run{13}, it is clear that the mass parameter is mostly constrained by the prior, then it appears that the small separation is the observable that strongly controls the estimates of the other parameters. 

The use of just $\als$ introduces a new problem. Studying the marginal PPD for $\stage$, it appears that most of the models lie close to the upper limit of the domain we defined. This means that our prior is very likely to be incorrect. However, when increasing the range for $\stage$, the MCMC algorithm does not converge anymore. This is seen by the fact that our 8 chains are stuck in very different regions of the space of parameters. 
Note that the lack of convergence when sampling a larger volume of $\pspace$ could be identified thanks to our use of several parallel Markov chains. Indeed, if considered individually, chains getting stuck in a local maximum could appear as the solution of our inverse problem. This shows the importance of a careful monitoring of the convergence of MCMC algorithm. This issue might be resolved with more sophisticated version of the MCMC algorithm \citep{Robert05,Gregory05,Benomar10,Handberg11}. 

\subsubsection{Structure of \acena}\label{sect:5D_struct}

\begin{figure*}
\begin{center}

\includegraphics[width=\textwidth]{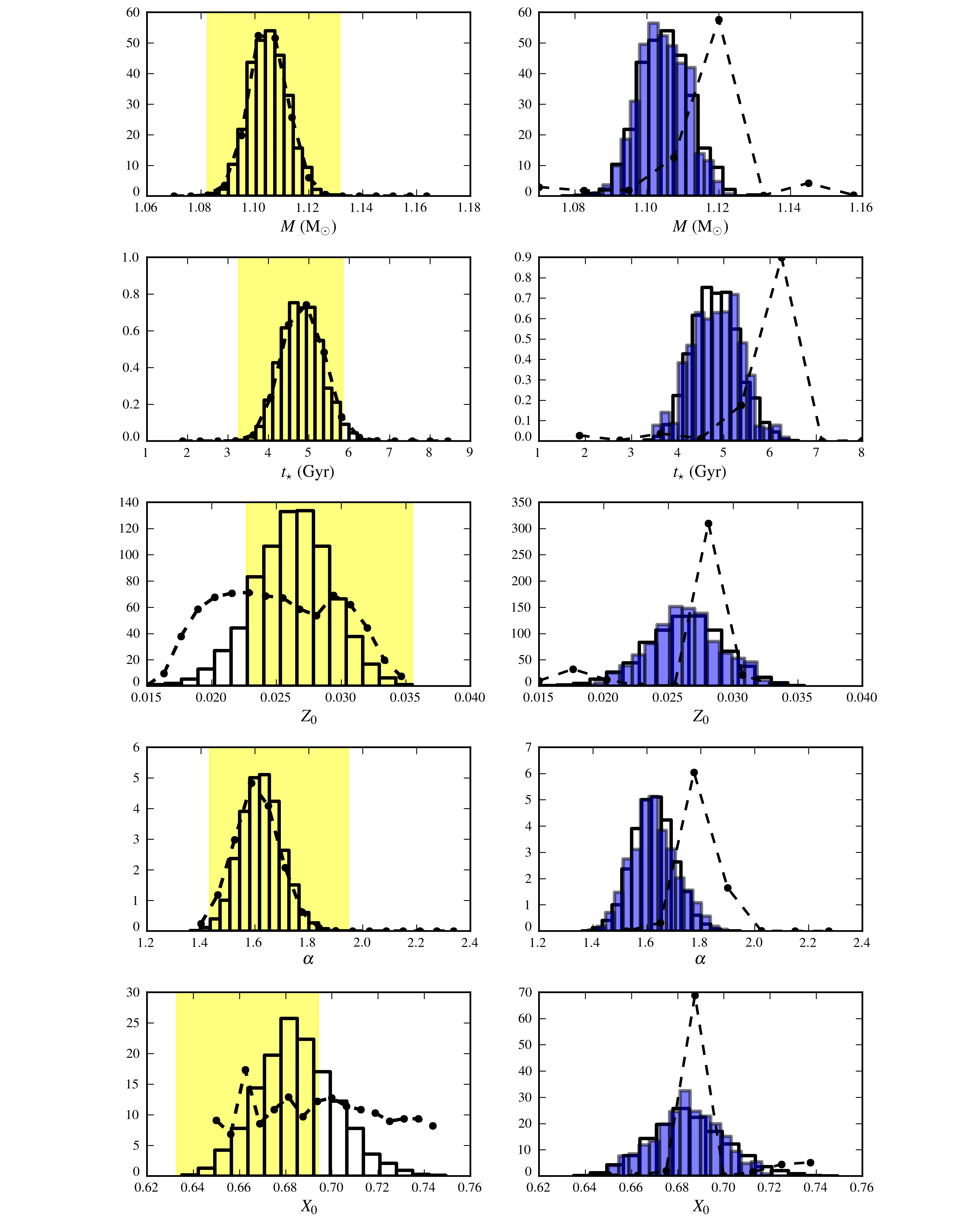}
\caption{\emph{Left column}: The histograms represent the marginal PPDs for $M$, $\stage$, $Z_0$, $\alpha$ and $X_0$ obtained from \run{13} ($\Xv = (\teff,L,R,\ass)$, Gaussian prior on the mass). The dots represent the same distribution but obtained from the integration of a grid of size $\sim16^5$. The yellow-shaded area mark the range for each parameter within which models with convective cores have been obtained. \emph{Right column}: The white histogram are the same as in left column. The blue histograms are the marginal PPDs but with a reduced Markov chain. The dots represent the same marginal PPDs but obtained from the integration of a subgrid of $\sim 8^5$ models}
\label{fig:mPPD_5D}
\end{center}
\end{figure*}
\begin{figure*}
\begin{center}
\includegraphics[width=\textwidth]{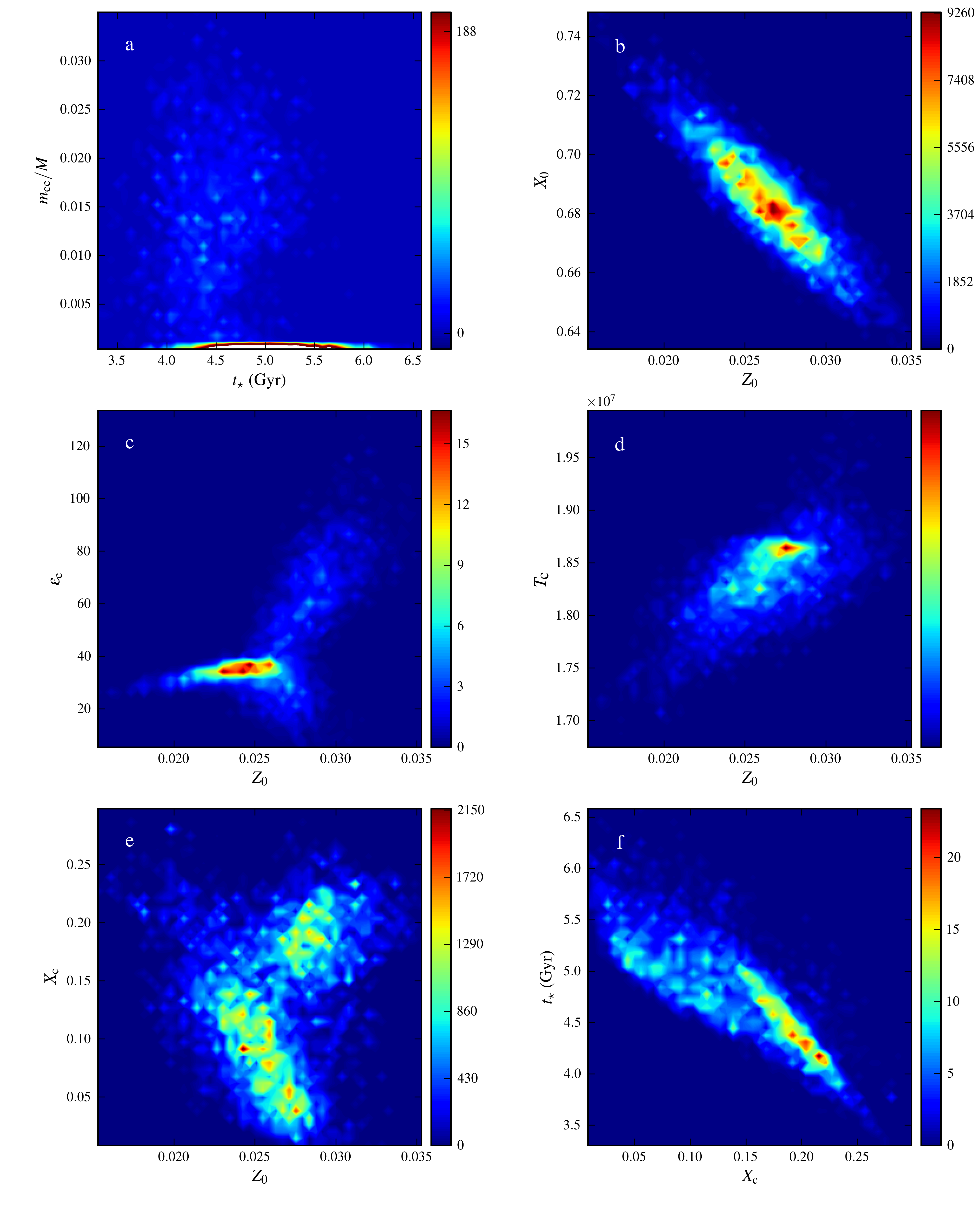}
\caption{2-dimensional marginal Posterior Probability Densities for a) the age of {\acena}, and relative mass of its convective core (note that the PPD has been truncated in order to obtain sufficient contrast where $\mcc > 0$), b) the initial metallicity and hydrogen mass fraction c) the initial metallicity and central energy production rate, d) the initial metallicity and the central temperature, e) the initial metallicity and central hydrogen mass fraction, f) the central hydrogen mass fraction and age.}
\label{fig:PPD2D_5D}
\end{center}
\end{figure*}

The picture of the stellar interior we obtain from the  MCMC simulations in the 5-dimensional parameter space is sensibly different from those deduced from the 2-dimensional exercise. This is particularly true when looking at the potential presence of the convective core. The mass prior and the atmosphere-dependent observables do not play such a critical role as they did in Sect~\ref{sect:2D_struct} in discarding the possibility of convection at the center of \acena. 

As before, we find a non-negligible probability of having a convective core when the only seismic observable included in $\Xv$ is the average small separation. We note that this probability might vary slightly from one chain to the other, but we still obtain an approximate probability of existence for the convective core 
$P(m_{\mathrm{cc}},r_{\mathrm{cc}} \neq 0|\Xv) \sim 0.40$. This is much higher than the value obtain in the 2-dimensional case. Adding the average large separation does not change this number the probability being in that case $\sim$42\%. We can again put an upper limit on the mass and radius of a possible convective core. For \run{12} we obtain $r_{cc} \lesssim 0.059$~R$_{\star}$  and $m_{cc} \lesssim 0.035$~M$_{\star}$, for \run{13} $r_{cc} \lesssim 0.055$~R$_{\star}$  and $m_{cc} \lesssim 0.030$~M$_{\star}$. The difference being small, it shows that it is really the use of a precise value of $\ass$ that controls the estimated characteristics of the convective core. In Fig.~\ref{fig:cc5D} we show the posterior probability densities obtained for $r_{cc}$ from runs \#12 and \#13. We see some small departures from gaussianity. If we consider the distribution for non-null radius and mass and compute the MAP and its associated credible set\footnote{In this case, we could not assume that the densities were Gaussian, therefore, we used the MAP as the estimate of the age and defined the $(1-\eta)$\% credible set as $C_k = \{\theta_k \in{\pspace}_k:\pi(\theta_k |\Xv) \geq q(\eta)\}$, with $q$ the smallest constant such as $P(C_k|\Xv) > 1-\eta$, and $\pspace_k$ the subspace of $\pspace$ corresponding to parameter $\theta_k$.}, we can conclude that an existing convective core in {\acena} would have  $r_{cc} = 0.044^{+0.007}_{-0.011}$~R$_{\star}$  and $m_{cc} = 0.014^{+0.005}_{-0.002}$~M$_{\star}$ from \run{12}. Again, this is very close to the values we obtain from \run{13}, $r_{cc} = 0.044^{+0.007}_{-0.08}$~R$_{\star}$ and $m_{cc} = 0.016^{+0.002}_{-0.003}$~M$_{\star}$ from.

In Fig.~\ref{fig:mPPD_5D}, we represent, for each individual parameters the range within which are observed convective cores in \run{13}. We see that they span the entire probability densities for the mass, the age and the mixing-length parameter. We can on the other hand determine lower and upper limits for, respectively, $Z_0$ and $X_0$ for the onset of convection. The resulting condition is $Z_0 \gtrsim 0.23$ and $X_0 \lesssim 0.69$. It is interesting to note that this upper limit for $X_0$ does not correspond to the value we adopted in the 2-dimensional case. Whereas in Sect.~\ref{sect:2D_struct} we obtain some models with $X_0 = 0.7$ and a convective core, it is never the case in the 5-dimensional case. This means that when these parameters are anticorrelated and that when the initial hydrogen fraction is as high as we assumed when we fixed it, then $Z_0$ is much lower than the value of 0.027 we assumed above. Therefore, the models with convective cores that we observed in both case do not recover the same regions of the parameter space $\pspace$. This shows how difficult it is to make \emph{a priori} assumptions on the stellar parameters and how misleading this can be in some cases.

The general picture obtained from the 5-dimensional modelling is that the critical parameters for the existence of a convective core are those controlling the chemical composition. Interestingly enough, the precise value of the mass is not relevant for the formation of these structures, in the sense that the distribution of the $m_{\mathrm{cc}}$ and $r_{\mathrm{cc}}$ for {\acena} is independent of $M$.
This is slightly surprising since the appearance of a convective core is linked to the transition from the pp cycle to the CNO cycle as the main source of energy production at the stellar centre. On the ZAMS, when the temperature is high enough, the energy production rate of the latter ($\ecno$) overcomes the one from the former ($\epp$) by several orders of magnitude. For a given chemical composition, $\ecno$ increases very rapidly with the temperature and even the modest rise of central temperature with mass is enough to trigger the burning of hydrogen through the CNO cycle. The substantial increase in the nuclear energy rate that results leads in turn to an increase of $\grad$, hence to convectively unstable layers. However, in our results, we observe no correlation between the mass of {\acena} and the existence (or size or mass) of a convective core. It is noteworthy that this result also holds for \run{12}. This means that, even though the absolute value of the mass is important (low-mass stars will almost always be in the solar or M-dwarf regimes, higher-mass stars will always have a convective core), at the precision level obtained on this parameter\footnote{And even in the case of stars other than {\acena} for which the presence of a convective core is dubious: precisions on the mass of the order of what we report in \run{12} are now routinely claimed from asteroseismic studies.}, it is not its potential variations that will affect the development of the inner convective instabilities.

We thus need to examine the influence of the other stellar parameters on the convective core formation process. As for the mass we observe them for all considered values of $\stage$ and $\alpha$. However, the situation is different for the two parameters. The mixing-length parameter is essentially decorrelated from the mass and radius of the convective core. For the age, we need to look more closely to the distribution $\pi(\stage,\mcc|\Xv)$, which is represented in Fig.~\ref{fig:PPD2D_5D}a. We can see that, for $\mcc > 0$, it is skewed towards older ages. One thus cannot only rely on the extremum age values at which convective cores are observed and given in Fig.~\ref{fig:mPPD_5D}, since the upper limit is only representative of a tail of the 2-dimensional distribution $\pi(\stage,\mcc|\Xv)$. We will discuss below the implication of this preliminary remark.

We now turn our attention towards $X_0$ and $Z_0$. It must first be noted that these parameters are strongly anti-correlated (see Fig.~\ref{fig:PPD2D_5D}b). Therefore, in the following discussion, the effect of an increase in $Z_0$ can be directly translated into a decrease in $X_0$. We should also recall that on the main sequence, since only hydrogen is burned and helium produced, and because we have neglected microscopic diffusion, the initial metallicity correspond to the final metallicity. The main physical variables governing the energy rate \citep[see e.g.,][]{Kippenhahn94} are the temperature, $T$, the density $\rho$, the hydrogen mass fraction, $X$, and, in the case of the CNO cycle, the CNO-element mass fraction $\xcno$ (when we refer to their central values, we label them with the subscript {\textquotedblleft}c{\textquotedblright}). It is our goal here to understand how they are affected by the initial chemical composition of {\acena}. It can be seen in Fig.~\ref{fig:PPD2D_5D}c that ultimately $Z_0$ is correlated to $\varepsilon_{\mathrm{c}}$ (which is $\eppc+\ecnoc$) and this is the relation we want to elucidate.

\input{Tables/grid5D.tex}

Fig.~\ref{fig:PPD2D_5D}c and \ref{fig:PPD2D_5D}d clearly show that two families of models exist for {\acena}. In the first one, the metallicity is comprised between approximately 0.020 and 0.026. In this interval, the nuclear energy production rate does not increase much whereas the central temperature is positively correlated with the metallicity. In the second regime, i.e. for $ 0.034 \gtrsim Z_0 \gtrsim 0.026$, $\varepsilon_{\textnormal{c}}$ increases with $Z_0$. On the contrary, $\tc$ does not seem to increase anymore and an absolute maximum in the marginal PPD is reached at temperatures $\sim 1.86\times10^7$~K.

We can interpret these two regimes as reflecting the energy generating cycle at play in the central layers of {\acena}. The first one correspond to the pp cycle. The increase of temperature with $Z_0$ correspond to an opacity effect in the higher layer of {\acena}. On the one hand, enhancing the metallicity increases the opacity in these regions\footnote{Most likely through an increase of H$^-$ opacity at the surface, which is sensitive to the metallicity. Other sources, like, bound-bound transitions or ionization of heavy atoms in deeper layers might also have an effect, albeit smaller.}. On the other, we observe that the surface luminosity is correlated with neither $Z_0$, $\ec$ nor $\tc$. Therefore, to produce the same surface luminosity the temperature near the core needs to increase so that $\ec$ does too. This raise in $\tc$ has relatively modest impact on the energy production rate, which is expected for the pp cycle.

Models of {\acena} with metallicities higher than $\gtrsim 0.026$ have clearly a different behaviour. We consider that they are stars for which the CNO cycle has already become the main source of nuclear energy in the central regions. The fact that the temperature does not increase with metallicity anymore is precisely due to the sensitivity of $\ecno$ to the temperature. Models with higher temperatures would have had much too large luminosities for {\acena}. Conversely, the fact that $\ec$ becomes strongly correlated with the initial metallicity can be explained by an increase in $\xcnoc$, on which the energy production rate depends linearly. Indeed, this quantity is considered to be proportional to $Z_0$ and hence $Z$. Since $\xcno$ can approximately double in the considered range for the metallicity, we are merely seeing the effect of this increase in the right part of the distribution in Fig.~\ref{fig:PPD2D_5D}c. It is extremely interesting to note that convective cores are present in {\acena} models almost immediately as the CNO cycle becomes dominant, the great concentration of the energy generation rate near the centre that triggering convection. Therefore, the CNO peaks of distributions in Fig.~\ref{fig:PPD2D_5D}c and \ref{fig:PPD2D_5D}d can be related straightforwardly to the peak for $\mcc > 0$ Fig.~\ref{fig:PPD2D_5D}a.

Finally, Fig.~\ref{fig:PPD2D_5D}e and \ref{fig:PPD2D_5D}f show that these two regimes also correspond to an age effect and we observe a clear bimodality in the distribution for $(Z_0,\xc)$. Models with initial metallicity $\gtrsim 0.026$ have a higher $\xc$ and are thus younger than those with lower metallicity. This is confirmed by Fig.~\ref{fig:PPD2D_5D}f, which, besides displaying the well-known age-central hydrogen fraction correlation also shows that in the CNO mode $\xc$ decreases less rapidly with age than in the pp regime. This is related to the presence of a convective core in these models, which allows the less hydrogen-depleted regions of the star to act like a reservoir for the centre. We can also note that in the CNO regime, $\xc$ seems to be positively correlated with $Z_0$. Therefore, there is a slight trend for models with a higher metallicity to be younger. This is clearly the opposite in the pp regime.

\subsubsection{Comparison with grid-based integration}\label{sect:grids5D}

The last step of our study was to compare the results from the MCMC simulations, in a similar fashion to what has been done in Sect.~\ref{sect:grids2D}, to the those obtained from integration on a grid of models. According to our preliminary remarks, we had to compute a sparser grid than what we did for the 2-dimensional case. We chose to sample the space of parameters using 16 points in each direction. This amounts to 914\,749 points in  $\pspace$, after discarding those which do not lead to a convergent solution using ASTEC.

In Fig.~\ref{fig:mPPD_5D}, we represented the marginal probability densities as obtained from direct integration based on our grid. The results are fairly similar to the MCMC simulation for the mass, the age and the mixing-length parameter. They differ significantly for $Z_0$ and $X_0$. Based on our discussion from Sect.~\ref{sect:grids2D}, we interpret this as a numerical effect stemming from the sparse sampling\footnote{But of course, for real data, one cannot rule out with absolute certitude the possibility that the MCMC algorithm got stuck in a local maximum and could not sample properly the PPD. However, given the fact that our 8 independent chains do agree, it seems highly unlikely here to be the case.}. Table~\ref{table:simu5Dgrid} also gives the first moment of the distribution obtained by the grid-integration procedure. Note that we chose to display them in the standard form of an estimate and the limits of an associated symmetrical credible interval. We did with the pedagogical purpose in mind to show that providing such statistical summaries without knowledge of the underlying density might be misleading. Even though the posterior standard deviations are close, one can see from Fig.~\ref{fig:mPPD_5D} that they do not recover the same reality of our state of knowledge. Therefore, one has to consider some of these values with care: the Gaussian approximation can very well be wrong, especially when dealing with the highly non-linear stellar evolutionary models. In this case of the grid-based determinations of $Z_0$ and $X_0$ the PM is not even the best estimator one can choose. One may rather want to use the MAP or the posterior median. The case for the metallicity is even more extreme. We see that the improper sampling leads to a multimodal distribution, and one might be tempted to distinguish two possible scenarios.

These departures between the MCMC and grid-sampling approach are, once more, of much lesser magnitude when the target PPD is broader. Therefore, we conclude that, since no predefine rule for a regular sampling of $\pspace$ exists, the MCMC approach is able, when the set of observable allows to constrain enough the PPD of the stellar parameters, to obtain more information on these densities, at an inferior computational cost. 

However, one might still be content with the results from the grid-based estimation and consider that they are close enough to what seems to be the correct PPD. Thus, the question we are now facing is to determine, even roughly, how much more efficient the MCMC could be in this 5-dimension parameter space and if this performance can be approached by a grid using the same number of model. To that effect, we selected a subgrid by rejecting one point out of two for each parameter. This leads us a total of $N \sim 8^5$ points sampling the volume of interest in $\pspace$. In Fig.~\ref{fig:mPPD_5D} (right column), we represented the marginal PPDs obtained from the integration of this subgrid and those from the MCMC simulation using a Markov chain of length equal to the number of points in the subgrid. The result is spectacular, in the sense that the grid approach fails severely to approximate the densities. On the other hand, the reduced MCMC sample is extremely close to the large one. This gives a convincing illustration that, for $\dim(\pspace) = 5$, the stochastic sample approach offered by the MCMC methodology is much more efficient than the grid strategy. Our conjecture is that this remains valid for $\dim(\pspace) > 5$.

It indeed is a well-known property of stochastic sampling methodologies that they tend to perform better than more systematic sampling schemes for high-dimensional parameter spaces. However, it was our goal to verify that this was already satisfied for stellar parameter spaces of dimension at least 5. We also wanted to verify if this could translate into significant computational gains. This example is a very good indication that it is the case and that stellar parameter estimation can benefit greatly from the use of MCMC strategies.

\subsection{Comparison with other studies}\label{sect:previous}

In this section we briefly compare our results to previous studies using seismic data for {\acena} \citep{Eggenberger04,Miglio05,deMeulenaer10}. It is important to first stress the methodological differences between these previous studies and ours. They both rely on optimization approaches, and none of them in a Bayesian framework. This poses obvious problems when comparing the results. First, it is not clear how fixing the mass compares with our use of a Gaussian prior. More importantly, without access to the PPDs of the parameters, it is not possible to assess whether or not the maximum values reported in these studies correspond to local or global extrema. We already saw in Sect.~\ref{sect:simu5D} that some PPDs were difficult to sample (in the 5-dimensional case, with $\Xv = (L,\teff,R,\als)$ and a Gaussian prior on the mass), therefore such a configuration might not be uncommon.

Different physics was also used in the stellar codes. The major point is that both studies used microscopic diffusion, which we have neglected. This should have an observable effect, but its nature remains difficult to anticipate. Gravitational settling will deplete the external convective zone of its heaviest elements. The initial metallicity might thus be increased, or the initial hydrogen mass fraction decreased, in order to reproduce the observed $Z/X$. The impact of this additional mechanism shall be studied with care in future studies using the MCMC approach.

Finally, {\acenb} has been modelled simultaneously in these studies. This introduces additional constraints on the solution to the inverse problem. We did not retain this strategy for logistical reasons. Modelling two stars at the same time would have implied significant modifications in our code. We thus preferred to focus on the star with the most reliable data (in particular the seismic data).

We should also consider the effect of the value they used for the average small separation, $\ass = 8.7 \pm 0.8$~$\mu$Hz used in these studies. To evaluate rapidly the impact this has on the estimate of the age, we can use the grid described in Sect.~\ref{sect:grids5D}. We obtain a relatively Gaussian marginal PPD and a lower age $\stage = 3.84 \pm 0.61$~$\mu$Hz for $\Xv = (L,\teff,R,\ass)$ and the Gaussian prior on the mass. Note that although we use the value provided by \citet{deMeulenaer10}, $\ass = 5.8 \pm 0.1$~$\mu$Hz we obtained $\stage = 6.04 \pm 0.35$~$\mu$Hz for the same $\Xv$ and prior. This agrees with the theoretical behaviour of the small separation, decreasing with age. We should however recall that although these grid-based are convenient because easily computed once the grid has been produced, we are still affected by the drawbacks we mentioned in Sect.~\ref{sect:grids5D}. The marginal PPDs for the $Z_0$, $\alpha$ and $X_0$ are obviously problematic and a proper MCMC simulation would be necessary (or a denser grid on larger scales) would be necessary to provide acceptable estimates of these parameters. 

Keeping in mind these remarks, we see that for the age of {\acena} we estimate in the 5-dimensional (runs \#12 and \#13) is significantly lower than those quoted in \citet{Miglio05}. Their A1e model seems to be the more relevant for a comparison with our own results. They obtain an age of $6.7\pm0.5$~Gyr. This is the main departure between the two studies. The other parameters are all within their respective error bars (which are, as a general rule commensurate). However, given the complex interplay between the stellar parameters and other variables such as, for instance, $\xc$, a change in one of them, even within the quoted error bars might very well lead to a very significant departure of another parameter. Only an homogeneous comparison using both models, with the same physical prescriptions, can give us more information on the differences between our studies. The same kind of picture can be drawn from a comparison between our work and the results of \citet{Eggenberger04}

The problem of the uncertainties is also difficult to tackle. As a general rule, those from \citet{Eggenberger04} seem to be underestimated with respect to ours. However, their method for estimating credible intervals is much less robust and statistically sound than what is allowed from the study of the marginal PPDs of the stellar parameters. \citet{Miglio05} used a Hessian-based approach and obtained uncertainties commensurate with ours. Our study also provides a confirmation of theirs since they need to make the implicit assumption that they are dealing with normally-distributed parameters before inverting their Hessian matrix. Note also that if this is not the case, this methodology is known to underestimate the uncertainties.

One could also contend that the differences come from the physics in our models, in particular the use of the EFF equation of state. This is unlikely since we see in \citet{Miglio05} that a change of equation of state only modestly affects the final estimates (see for instance their models A1e and A1r). Therefore, we suggest that most of the difference should be attributed to the adopted value for $\ass$.

Finally, we note that gyrochronology allows to obtain some measurements of the age of the stars. One should note that these could have been used as a prior on the age. We chose not to do so, in particular because we preferred to constrain the age using seismic quantities rather than by the use of gyrochronology. Nevertheless, we note that our estimates using the average small separations give an age in agreement with the results from \citet{Barnes07}. On the other hand, the estimates from \citet{Delorme11} are much higher and agree more closely with our estimates using $\als$. It is an interesting fact that the same technique provide such different estimates.

%% file: Tables/grid5D.tex
\begin{table*}
\begin{center}
\caption{Same as Table~\ref{table:simu5D} but with estimates obtained from the grid described in Sect.~\ref{sect:grids5D}.}
\label{table:simu5Dgrid}
\begin{tabular}{lcccccccc}
\toprule %\hline

Run \#&$\Xv$&$\pi(M)$& $\hat{M}$ ({\msol})& $\hat{\stage}$ (Gyr) & $Z_0$ & $\alpha$ & $X_0$\\

\midrule

12&$[L, \teff, R, \langle \delta \nu \rangle]$                            &Uniform  &$1.12\pm 0.03$ &$4.9\pm0.5$ &$0.026\pm0.005$ &$1.6\pm0.1$  &$0.70\pm0.02$ \\

13&$[L, \teff, R, \langle \delta \nu \rangle]$                            &Gaussian &$1.105\pm0.007$ &$4.9\pm0.5$ &$0.025\pm0.005$  &$1.61\pm0.08$ &$0.70\pm0.03$ \\

14&$[L, \teff, R, \langle \delta \nu \rangle, \langle \Delta \nu \rangle]$&Gaussian& $1.102\pm0.005$ & $5.0\pm0.5$ & $0.024\pm0.004$ & $1.58 \pm 0.08$ & $0.70\pm0.02$ \\

\bottomrule
\end{tabular}
\end{center}
\end{table*}

%% file: Sections/Conclusion.tex
In this paper, we applied a Bayesian methodology to solve the inverse problem that consists in estimating the stellar parameters of {\acena}, some observable quantities being known. To that purpose, we used an MCMC algorithm in its classical Metropolis-Hastings form. We repeated the experiment twice, first fitting the data by varying only the mass and the age, and using assumptions to fix the mixing-length parameter and the initial hydrogen fraction and metallicity; second letting all five parameters free.

In each case we were able to obtain the posterior probability densities for all considered parameters and their derived values, in particular the mass and radius of the convective core. We observed that different families of solutions emerged depending whether the observations we used are sensitive to the innermost regions of the star (the average small separation) or to its outer layers (all the others). This remarkable fact allows us in particular to reopen the case for a presence of a convective core in {\acena}. Expressed in probabilistic terms, we obtained non-negligible odds for the existence of such a structure, sometimes reaching over 40\%.

From a numerical standpoint, we also produced evidence that the MCMC approach offers a good and efficient alternative to methods generally used in stellar physics, either optimization or grid-based Bayesian approaches. More precisely, when the dimension of the parameter space increases, it becomes less computationally expensive than the straightforward grid integration approach (assuming one wishes to obtain the same statistical output). It also preserves the advantage of Bayesian methodologies with respect to optimization methods in the sense that we have access to the full posterior probability density. Considering that it is already efficient and that there exist other variants of the MH algorithm that can improve the efficiency of our own, we deem this method as very interesting for the future of stellar parameter estimation.

Finally, it seems clear that further progress on the understanding of the inner structure of {\acena} will require more precise and accurate data. It is in particular our hope that asteroseismology can help to unravel the convective core puzzle. We have already seen that precise values of $\ass$ affect greatly our answer to this problem. For this, one will need very long and precise radial velocity measurements, if possible continuous. The MCMC methodology will then be of great use to provide robust estimates of the stellar parameters and their uncertainties, knowing these new data.

%% file: Appendices/Conv.tex
\subsection{Algorithm monitoring}\label{sect:conv}

Assessing the convergence of an MCMC algorithm to its target distribution is a vast and complex problem. There is no clear-cut answer to this issue and most of the current methods are empirical ones. We present in this section the few simple indicators we used to monitor convergence in the present study. 

\subsubsection{Burn-in sequence}

As seen in Algorithm~\ref{algo:MHalg}, the initialization of the process involves the input of an initial value for the parameter. Hence, during a first phase, traditionally called the burn-in, the random generation will not be done according to the target PPD. It is then necessary to discard the first iterations of the Markov chain in order to yield the {\it stationary} regime. It is difficult to define with precision the length of the burn-in sequence, which obviously depends on the target distribution and of the initial guess.

For an MCMC sample to approximate the targeted PPD, it is necessary that the $\thetav^{(t)}$ are generated randomly\footnote{Not sufficient though: one has still to determine if they are indeed generated according to the desired PPD.} and, hence, are independent of each other.

To verify this, one can compute the autocorrelation function (ACF), which can be written, for a Markov Chain of length $N$
\begin{equation}\label{ACF}
\rho^{\theta_i}(k)=\frac{\displaystyle\sum^{N-(k+1)}_{k}(\theta_i^{(t)}-\bar{\theta})(\theta^{(t+k)}_{i}-\bar{\theta})}{\sqrt{\displaystyle\sum^{N-(k+1)}_{k}(\theta_i^{(t)}-\bar{\theta})^2}\sqrt{\displaystyle\sum^{N-(k+1)}_{k}(\theta^{(t+k)}_{i}-\bar{\theta})^2}},
\end{equation}

the denominator being a normalization factor. A value close to zero is a good indication that the values of the sequence have indeed been generated randomly. The behaviour of the ACF is grossly that of a decreasing exponential $\exp{(-k/\ell_{\rho})}$, $\ell_{\rho}$ being the characteristic length  for decorrelation \citep{Gregory05}. 

For each run, we tested burn-in sequences of size $\sim\ell_{\rho}$ and higher. Usually, discarding a number of iteration twice as large as $\ell_{\rho}$ guarantees that the sample is indeed stationarily generated.

\subsubsection{Convergence testing}

As mentioned above, MCMC algorithms rely on the convergence properties of the Markov Chains. Therefore, the number of iterations has to be large enough in order to guarantee that the Markov Chain asymptotic regime is a good approximation to the target PPD. 
Assuming that an adequate burn-in sequence has been removed, the question arises of the stopping rule.

A classical way to address the problem of convergence is to use empirical graphical indicators. It is particularly interesting to follow the evolution of the empirical cumulative average estimator
\begin{equation}{\label{eq:cumav}}
S_T=\frac{1}{T}\displaystyle\sum_{t=1}^{T} h(\thetav^{(t)}),
\end{equation}
where $h$ is an arbitrary function of $\thetav$  and $T$ the length of the Markov Chain. In practice, we will be mostly interested by the function $h(\thetav) = \theta_i$ for $i=1,\dots,n_p$, so that $S_T$ approximates the Posterior Mean estimate~(\ref{eq:pm}). 
At convergence, variations in $S_T$ (as a function of $T$) should be negligible.

Depending of the dimension of the parameter space we considered, we ran $M=3$ or $M=8$ chains with different initializations. This was done in order to explore fully the space of parameter and eventual dependence on the initial guess.
In this configuration, instead of $S_T$, we consider the average over the $M$ chains:
\begin{equation}\label{mult_mean}
\langle h(\thetav) \rangle = \displaystyle \frac{1}{M}\sum^M_{m=1} \frac{1}{T}\sum^T_{t=1} h(\thetav^{(t)}_m),
\end{equation}
where $m$ indexes the number of the chain, and the associated average variance (\emph{within-chain variance})
\begin{equation}\label{within}
\displaystyle W_T = \frac{1}{M} \sum_{m=1}^{M}\frac{1}{T-1}\sum_{t=1}^{T}\left(\theta_m^{(t)} - \frac{1}{T}\sum_{t=1}^{T}\theta_{m}^{(t)}\right)^2.
\end{equation}

One can also define the \emph{between-chain variance}
\begin{equation}\label{within}
B_T = \frac{1}{M-1}\sum_{m=1}^{M}\left(\frac{1}{T}\sum_{t=1}^{T}\theta_{m}^{(t)}-\frac{1}{M}\sum_{m=1}^{M}\frac{1}{T}\sum_{t=1}^{T}\theta_m^{(t)}\right)^2
\end{equation}
this other indicator can be used to check convergence as explained in \citet{Gelman95}.

%% file: acena_MCMCI.bbl
\begin{thebibliography}{79}
\expandafter\ifx\csname natexlab\endcsname\relax\def\natexlab#1{#1}\fi

\bibitem[{{Andrieu} \& {Doucet}(1999)}]{Andrieu99}
{Andrieu} C., {Doucet} A., 1999, IEEE Transactions on Signal Processing, 47,
  2667

\bibitem[{{Angulo} {et~al}\mbox{.}(1999){Angulo}, {Arnould}, {Rayet},
  {Descouvemont}, {Baye}, {Leclercq-Willain}, {Coc}, {Barhoumi}, {Aguer},
  {Rolfs}, {Kunz}, {Hammer}, {Mayer}, {Paradellis}, {Kossionides}, {Chronidou},
  {Spyrou}, {degl'Innocenti}, {Fiorentini}, {Ricci}, {Zavatarelli},
  {Providencia}, {Wolters}, {Soares}, {Grama}, {Rahighi}, {Shotter}, \& {Lamehi
  Rachti}}]{Angulo99}
{Angulo} C. {et~al.}, 1999, Nuclear Physics A, 656, 3

\bibitem[{{Antia} \& {Basu}(2005)}]{Antia05}
{Antia} H.~M., {Basu} S., 2005, \apjl, 620, L129

\bibitem[{{Asplund} {et~al}\mbox{.}(2004){Asplund}, {Grevesse}, {Sauval},
  {Allende Prieto}, \& {Kiselman}}]{Asplund04}
{Asplund} M., {Grevesse} N., {Sauval} A.~J., {Allende Prieto} C., {Kiselman}
  D., 2004, \aap, 417, 751

\bibitem[{{Asplund} {et~al}\mbox{.}(2009){Asplund}, {Grevesse}, {Sauval}, \&
  {Scott}}]{Asplund09}
{Asplund} M., {Grevesse} N., {Sauval} A.~J., {Scott} P., 2009, \araa, 47, 481

\bibitem[{{Barnes}(2007)}]{Barnes07}
{Barnes} S.~A., 2007, \apj, 669, 1167

\bibitem[{{Bazot} {et~al}\mbox{.}(2007){Bazot}, {Bouchy}, {Kjeldsen},
  {Charpinet}, {Laymand}, \& {Vauclair}}]{Bazot07}
{Bazot} M., {Bouchy} F., {Kjeldsen} H., {Charpinet} S., {Laymand} M.,
  {Vauclair} S., 2007, \aap, 470, 295

\bibitem[{{Bazot}, {Bourguignon} \& {Christensen-Dalsgaard}(2008){Bazot},
  {Bourguignon}, \& {Christensen-Dalsgaard}}]{Bazot08}
{Bazot} M., {Bourguignon} S., {Christensen-Dalsgaard} J., 2008, \memsai, 79,
  660

\bibitem[{{Bazot} {et~al}\mbox{.}(2012){Bazot}, {Campante}, {Chaplin},
  {Carfantan}, {Bedding}, {Dumusque}, {Broomhall}, {Th{\'e}ado}, {van Grootel},
  {Arentoft}, {Asplund}, {Castro}, {Christensen-Dalsgaard}, {Do Nascimento},
  {Dintrans}, {Kjeldsen}, {Monteiro}, {Santos}, {Sousa}, \&
  {Vauclair}}]{Bazot12}
{Bazot} M. {et~al.}, 2012, \aap

\bibitem[{{Bazot} {et~al}\mbox{.}(2011){Bazot}, {Ireland}, {Huber}, {Bedding},
  {Broomhall}, {Campante}, {Carfantan}, {Chaplin}, {Elsworth}, {Mel{\'e}ndez},
  {Petit}, {Th{\'e}ado}, {van Grootel}, {Arentoft}, {Asplund}, {Castro},
  {Christensen-Dalsgaard}, {Do Nascimento}, {Dintrans}, {Dumusque}, {Kjeldsen},
  {McAlister}, {Metcalfe}, {Monteiro}, {Santos}, {Sousa}, {Sturmann},
  {Sturmann}, {Ten Brummelaar}, {Turner}, \& {Vauclair}}]{Bazot11}
---, 2011, \aap, 526, L4

\bibitem[{{Bedding} {et~al}\mbox{.}(2004){Bedding}, {Kjeldsen}, {Butler},
  {McCarthy}, {Marcy}, {O'Toole}, {Tinney}, \& {Wright}}]{Bedding04}
{Bedding} T.~R., {Kjeldsen} H., {Butler} R.~P., {McCarthy} C., {Marcy} G.~W.,
  {O'Toole} S.~J., {Tinney} C.~G., {Wright} J.~T., 2004, \apj, 614, 380

\bibitem[{{Benomar} {et~al}\mbox{.}(2010){Benomar}, {Baudin}, {Marques},
  {Goupil}, {Lebreton}, \& {Deheuvels}}]{Benomar10}
{Benomar} O., {Baudin} F., {Marques} J.~P., {Goupil} M.~J., {Lebreton} Y.,
  {Deheuvels} S., 2010, Astronomische Nachrichten, 331, 956

\bibitem[{{B{\"o}hm-Vitense}(1958)}]{BV58}
{B{\"o}hm-Vitense} E., 1958, Zeitschrift fur Astrophysik, 46, 108

\bibitem[{{Bouchy} \& {Carrier}(2002)}]{Bouchy02}
{Bouchy} F., {Carrier} F., 2002, \aap, 390, 205

\bibitem[{{Brewer} {et~al}\mbox{.}(2007){Brewer}, {Bedding}, {Kjeldsen}, \&
  {Stello}}]{Brewer07}
{Brewer} B.~J., {Bedding} T.~R., {Kjeldsen} H., {Stello} D., 2007, \apj, 654,
  551

\bibitem[{{Brown} \& {Christensen-Dalsgaard}(1998)}]{Brown98}
{Brown} T.~M., {Christensen-Dalsgaard} J., 1998, \apjl, 500, L195+

\bibitem[{{Brown} {et~al}\mbox{.}(1994){Brown}, {Christensen-Dalsgaard},
  {Weibel-Mihalas}, \& {Gilliland}}]{Brown94}
{Brown} T.~M., {Christensen-Dalsgaard} J., {Weibel-Mihalas} B., {Gilliland}
  R.~L., 1994, \apj, 427, 1013

\bibitem[{{Caffau} {et~al}\mbox{.}(2008){Caffau}, {Ludwig}, {Steffen}, {Ayres},
  {Bonifacio}, {Cayrel}, {Freytag}, \& {Plez}}]{Caffau08}
{Caffau} E., {Ludwig} H.-G., {Steffen} M., {Ayres} T.~R., {Bonifacio} P.,
  {Cayrel} R., {Freytag} B., {Plez} B., 2008, \aap, 488, 1031

\bibitem[{{Castro}, {Vauclair} \& {Richard}(2007){Castro}, {Vauclair}, \&
  {Richard}}]{Castro07}
{Castro} M., {Vauclair} S., {Richard} O., 2007, \aap, 463, 755

\bibitem[{{Chmielewski} {et~al}\mbox{.}(1992){Chmielewski}, {Friel}, {Cayrel de
  Strobel}, \& {Bentolila}}]{Chmielewski92}
{Chmielewski} Y., {Friel} E., {Cayrel de Strobel} G., {Bentolila} C., 1992,
  \aap, 263, 219

\bibitem[{{Christensen-Dalsgaard}(1982)}]{JCD82b}
{Christensen-Dalsgaard} J., 1982, \mnras, 199, 735

\bibitem[{{Christensen-Dalsgaard}(1993)}]{JCD93}
---, 1993, in Astronomical Society of the Pacific Conference Series, Vol.~42,
  GONG 1992. Seismic Investigation of the Sun and Stars, {Brown} T.~M., ed., p.
  347

\bibitem[{{Christensen-Dalsgaard}(2008{\natexlab{a}})}]{JCD08b}
---, 2008{\natexlab{a}}, \apss, 316, 113

\bibitem[{{Christensen-Dalsgaard}(2008{\natexlab{b}})}]{JCD08a}
---, 2008{\natexlab{b}}, \apss, 316, 13

\bibitem[{{da Silva} {et~al}\mbox{.}(2006){da Silva}, {Girardi}, {Pasquini},
  {Setiawan}, {von der L{\"u}he}, {de Medeiros}, {Hatzes}, {D{\"o}llinger}, \&
  {Weiss}}]{daSilva06}
{da Silva} L. {et~al.}, 2006, \aap, 458, 609

\bibitem[{{de Meulenaer} {et~al}\mbox{.}(2010){de Meulenaer}, {Carrier},
  {Miglio}, {Bedding}, {Campante}, {Eggenberger}, {Kjeldsen}, \&
  {Montalb{\'a}n}}]{deMeulenaer10}
{de Meulenaer} P., {Carrier} F., {Miglio} A., {Bedding} T.~R., {Campante}
  T.~L., {Eggenberger} P., {Kjeldsen} H., {Montalb{\'a}n} J., 2010, \aap, 523,
  A54

\bibitem[{{Delorme} {et~al}\mbox{.}(2011){Delorme}, {Collier Cameron}, {Hebb},
  {Rostron}, {Lister}, {Norton}, {Pollacco}, \& {West}}]{Delorme11}
{Delorme} P., {Collier Cameron} A., {Hebb} L., {Rostron} J., {Lister} T.~A.,
  {Norton} A.~J., {Pollacco} D., {West} R.~G., 2011, \mnras, 413, 2218

\bibitem[{{Deubner} \& {Gough}(1984)}]{Deubner84}
{Deubner} F.-L., {Gough} D., 1984, \araa, 22, 593

\bibitem[{{Eggenberger} {et~al}\mbox{.}(2004){Eggenberger}, {Charbonnel},
  {Talon}, {Meynet}, {Maeder}, {Carrier}, \& {Bourban}}]{Eggenberger04}
{Eggenberger} P., {Charbonnel} C., {Talon} S., {Meynet} G., {Maeder} A.,
  {Carrier} F., {Bourban} G., 2004, \aap, 417, 235

\bibitem[{{Eggleton}, {Faulkner} \& {Flannery}(1973){Eggleton}, {Faulkner}, \&
  {Flannery}}]{EFF}
{Eggleton} P.~P., {Faulkner} J., {Flannery} B.~P., 1973, \aap, 23, 325

\bibitem[{{Endl} {et~al}\mbox{.}(2001){Endl}, {K{\"u}rster}, {Els}, {Hatzes},
  \& {Cochran}}]{Endl01}
{Endl} M., {K{\"u}rster} M., {Els} S., {Hatzes} A.~P., {Cochran} W.~D., 2001,
  \aap, 374, 675

\bibitem[{{Gabriel}(1989)}]{Gabriel89}
{Gabriel} M., 1989, \aap, 226, 278

\bibitem[{Gelman(1996)}]{Gelman95}
Gelman A., 1996, Boca Raton, pp. 131--143

\bibitem[{{Girardi} {et~al}\mbox{.}(2000){Girardi}, {Bressan}, {Bertelli}, \&
  {Chiosi}}]{Girardi00}
{Girardi} L., {Bressan} A., {Bertelli} G., {Chiosi} C., 2000, \aaps, 141, 371

\bibitem[{{Gough}(1986)}]{Gough86}
{Gough} D.~O., 1986, in Hydrodynamic and Magnetodynamic Problems in the Sun and
  Stars, {Osaki} Y., ed., p. 117

\bibitem[{{Grec}, {Fossat} \& {Pomerantz}(1983){Grec}, {Fossat}, \&
  {Pomerantz}}]{Grec83}
{Grec} G., {Fossat} E., {Pomerantz} M.~A., 1983, \solphys, 82, 55

\bibitem[{{Gregory}(2005)}]{Gregory05}
{Gregory} P.~C., 2005, {Bayesian Logical Data Analysis for the Physical
  Sciences: A Comparative Approach with `Mathematica' Support}. Cambridge
  University Press

\bibitem[{{Grevesse}, {Noels} \& {Sauval}(1993){Grevesse}, {Noels}, \&
  {Sauval}}]{Grevesse93}
{Grevesse} N., {Noels} A., {Sauval} A.~J., 1993, \aap, 271, 587

\bibitem[{{Gruberbauer}, {Guenther} \& {Kallinger}(2012){Gruberbauer},
  {Guenther}, \& {Kallinger}}]{Gruberbauer12}
{Gruberbauer} M., {Guenther} D.~B., {Kallinger} T., 2012, \apj, 749, 109

\bibitem[{{Guenther} \& {Demarque}(2000)}]{Guenther00}
{Guenther} D.~B., {Demarque} P., 2000, \apj, 531, 503

\bibitem[{{Guzik}, {Watson} \& {Cox}(2005){Guzik}, {Watson}, \&
  {Cox}}]{Guzik05}
{Guzik} J.~A., {Watson} L.~S., {Cox} A.~N., 2005, \apj, 627, 1049

\bibitem[{{Handberg} \& {Campante}(2011)}]{Handberg11}
{Handberg} R., {Campante} T.~L., 2011, \aap, 527, A56

\bibitem[{{Hastings}(1970)}]{Hastings70}
{Hastings} W.~K., 1970, Biometrika, 57, 97

\bibitem[{Idier(2008)}]{Idier08}
Idier J., ed., 2008, Bayesian Approach to Inverse Problems. ISTE Ltd and John
  Wiley \& Sons Inc

\bibitem[{{Iglesias} \& {Rogers}(1996)}]{Iglesias96}
{Iglesias} C.~A., {Rogers} F.~J., 1996, \apj, 464, 943

\bibitem[{{J{\o}rgensen} \& {Lindegren}(2005)}]{Jorgensen05}
{J{\o}rgensen} B.~R., {Lindegren} L., 2005, \aap, 436, 127

\bibitem[{{Kervella} {et~al}\mbox{.}(2003){Kervella}, {Th{\'e}venin},
  {S{\'e}gransan}, {Berthomieu}, {Lopez}, {Morel}, \& {Provost}}]{Kervella03}
{Kervella} P., {Th{\'e}venin} F., {S{\'e}gransan} D., {Berthomieu} G., {Lopez}
  B., {Morel} P., {Provost} J., 2003, {\aap}, 404, 1087

\bibitem[{{Kippenhahn} \& {Weigert}(1994)}]{Kippenhahn94}
{Kippenhahn} R., {Weigert} A., 1994, {Stellar Structure and Evolution}

\bibitem[{{Kjeldsen}, {Bedding} \& {Christensen-Dalsgaard}(2008){Kjeldsen},
  {Bedding}, \& {Christensen-Dalsgaard}}]{Kjeldsen08}
{Kjeldsen} H., {Bedding} T.~R., {Christensen-Dalsgaard} J., 2008, \apjl, 683,
  L175

\bibitem[{{Metcalfe}(2003)}]{Metcalfe03}
{Metcalfe} T.~S., 2003, \apss, 284, 141

\bibitem[{{Metropolis}(1953)}]{Metropolis53}
{Metropolis} N., 1953, \jcp, 21, 1087

\bibitem[{{Miglio} \& {Montalb{\'a}n}(2005)}]{Miglio05}
{Miglio} A., {Montalb{\'a}n} J., 2005, \aap, 441, 615

\bibitem[{{Montalb{\'a}n} \& {Miglio}(2006)}]{Montalban06}
{Montalb{\'a}n} J., {Miglio} A., 2006, Memorie della Societa Astronomica
  Italiana, 77, 472

\bibitem[{{Montalb{\'a}n} {et~al}\mbox{.}(2004){Montalb{\'a}n}, {Miglio},
  {Noels}, {Grevesse}, \& {di Mauro}}]{Montalban04}
{Montalb{\'a}n} J., {Miglio} A., {Noels} A., {Grevesse} N., {di Mauro} M.~P.,
  2004, in ESA Special Publication, Vol. 559, SOHO 14 Helio- and
  Asteroseismology: Towards a Golden Future, {Danesy} D., ed., p. 574

\bibitem[{{Morel}(1997)}]{Morel97}
{Morel} P., 1997, \aaps, 124, 597

\bibitem[{{Morel} {et~al}\mbox{.}(2000){Morel}, {Provost}, {Lebreton},
  {Th{\'e}venin}, \& {Berthomieu}}]{Morel00}
{Morel} P., {Provost} J., {Lebreton} Y., {Th{\'e}venin} F., {Berthomieu} G.,
  2000, \aap, 363, 675

\bibitem[{{Murdoch} \& {Hearnshaw}(1993)}]{Murdoch93}
{Murdoch} K., {Hearnshaw} J.~B., 1993, The Observatory, 113, 79

\bibitem[{{Neuforge-Verheecke} \& {Magain}(1997)}]{Neuforge97}
{Neuforge-Verheecke} C., {Magain} P., 1997, \aap, 328, 261

\bibitem[{{Pietrinferni} {et~al}\mbox{.}(2004){Pietrinferni}, {Cassisi},
  {Salaris}, \& {Castelli}}]{Pietrinferni04}
{Pietrinferni} A., {Cassisi} S., {Salaris} M., {Castelli} F., 2004, \apj, 612,
  168

\bibitem[{{Pourbaix} {et~al}\mbox{.}(2002){Pourbaix}, {Nidever}, {McCarthy},
  {Butler}, {Tinney}, {Marcy}, {Jones}, {Penny}, {Carter}, {Bouchy}, {Pepe},
  {Hearnshaw}, {Skuljan}, {Ramm}, \& {Kent}}]{Pourbaix02}
{Pourbaix} D. {et~al.}, 2002, \aap, 386, 280

\bibitem[{{Quirion}, {Christensen-Dalsgaard} \& {Arentoft}(2010){Quirion},
  {Christensen-Dalsgaard}, \& {Arentoft}}]{Quirion10}
{Quirion} P.-O., {Christensen-Dalsgaard} J., {Arentoft} T., 2010, \apj, 725,
  2176

\bibitem[{Robert \& Casella(2005)}]{Robert05}
Robert C.~P., Casella G., 2005, Monte Carlo Statistical Methods (Springer Texts
  in Statistics). Springer-Verlag New York, Inc., Secaucus, NJ, USA

\bibitem[{{Rogers} \& {Nayfonov}(2002)}]{OPAL02}
{Rogers} F.~J., {Nayfonov} A., 2002, \apj, 576, 1064

\bibitem[{{Rogers}, {Swenson} \& {Iglesias}(1996){Rogers}, {Swenson}, \&
  {Iglesias}}]{OPAL96}
{Rogers} F.~J., {Swenson} F.~J., {Iglesias} C.~A., 1996, \apj, 456, 902

\bibitem[{{Roxburgh} \& {Vorontsov}(1994)}]{RV94}
{Roxburgh} I.~W., {Vorontsov} S.~V., 1994, \mnras, 267, 297

\bibitem[{{Schaller} {et~al}\mbox{.}(1992){Schaller}, {Schaerer}, {Meynet}, \&
  {Maeder}}]{Schaller96}
{Schaller} G., {Schaerer} D., {Meynet} G., {Maeder} A., 1992, \aaps, 96, 269

\bibitem[{{S{\"o}derhjelm}(1999)}]{Soderhjelm99}
{S{\"o}derhjelm} S., 1999, \aap, 341, 121

\bibitem[{{Stahn} \& {Gizon}(2008)}]{Stahn08}
{Stahn} T., {Gizon} L., 2008, \solphys, 251, 31

\bibitem[{{Takeda} {et~al}\mbox{.}(2007){Takeda}, {Ford}, {Sills}, {Rasio},
  {Fischer}, \& {Valenti}}]{Takeda07}
{Takeda} G., {Ford} E.~B., {Sills} A., {Rasio} F.~A., {Fischer} D.~A.,
  {Valenti} J.~A., 2007, \apjs, 168, 297

\bibitem[{{Tassoul}(1980)}]{Tassoul80}
{Tassoul} M., 1980, \apjs, 43, 469

\bibitem[{{Th{\'e}venin} {et~al}\mbox{.}(2002){Th{\'e}venin}, {Provost},
  {Morel}, {Berthomieu}, {Bouchy}, \& {Carrier}}]{Thevenin02}
{Th{\'e}venin} F., {Provost} J., {Morel} P., {Berthomieu} G., {Bouchy} F.,
  {Carrier} F., 2002, \aap, 392, L9

\bibitem[{{Thoul} {et~al}\mbox{.}(2003){Thoul}, {Scuflaire}, {Noels},
  {Vatovez}, {Briquet}, {Dupret}, \& {Montalban}}]{Thoul03}
{Thoul} A., {Scuflaire} R., {Noels} A., {Vatovez} B., {Briquet} M., {Dupret}
  M.-A., {Montalban} J., 2003, \aap, 402, 293

\bibitem[{{Trampedach}(2007)}]{Trampedach07}
{Trampedach} R., 2007, in American Institute of Physics Conference Series, Vol.
  948, Unsolved Problems in Stellar Physics: A Conference in Honor of Douglas
  Gough, {Stancliffe} R.~J., {Houdek} G., {Martin} R.~G., {Tout} C.~A., eds.,
  pp. 141--148

\bibitem[{{Turck-Chi{\`e}ze} {et~al}\mbox{.}(2004){Turck-Chi{\`e}ze},
  {Couvidat}, {Piau}, {Ferguson}, {Lambert}, {Ballot}, {Garc{\'{\i}}a}, \&
  {Nghiem}}]{T-C04}
{Turck-Chi{\`e}ze} S., {Couvidat} S., {Piau} L., {Ferguson} J., {Lambert} P.,
  {Ballot} J., {Garc{\'{\i}}a} R.~A., {Nghiem} P., 2004, Physical Review
  Letters, 93, 211102

\bibitem[{{Turcotte} \& {Christensen-Dalsgaard}(1998)}]{Turcotte98}
{Turcotte} S., {Christensen-Dalsgaard} J., 1998, in ESA Special Publication,
  Vol. 418, Structure and Dynamics of the Interior of the Sun and Sun-like
  Stars, {S.~Korzennik}, ed., p. 407

\bibitem[{{Van Hoolst} \& {Smeyers}(1991)}]{VanHoolst91}
{Van Hoolst} T., {Smeyers} P., 1991, \aap, 248, 647

\bibitem[{{Vandakurov}(1967)}]{Vandakurov67}
{Vandakurov} Y.~V., 1967, \azh, 44, 786

\bibitem[{{Vorontsov}(1991)}]{Vorontsov91}
{Vorontsov} S.~V., 1991, Soviet Astronomy, 35, 400

\bibitem[{{Vorontsov} \& {Zharkov}(1989)}]{Vorontsov89}
{Vorontsov} S.~V., {Zharkov} V.~N., 1989, Astrophysics and Space Physics
  Reviews, 7, 1

\end{thebibliography}
